\documentclass[journal, twoside]{IEEEtran}
\usepackage{subcaption}
\usepackage{indentfirst}
\usepackage{graphicx}
\usepackage{amssymb}
\usepackage{amsfonts}
\usepackage{mathrsfs}
\usepackage{leftidx}
\usepackage{color}
\usepackage{amsmath}
\usepackage{arydshln}
\usepackage{amsthm}
\usepackage{ragged2e}
\usepackage{cite}
\usepackage{enumerate}
\usepackage{longtable}
\usepackage{float}
\usepackage{stfloats}
\usepackage{hyperref}
\usepackage{algpseudocode}
\usepackage{algorithm}
\usepackage{makecell}
\usepackage{url}
\usepackage{enumitem}
\usepackage{multirow} 
\usepackage{booktabs}
\theoremstyle{plain}

\theoremstyle{remark}
\newtheorem{rem}{Remark}

\usepackage{caption}
\usepackage{makecell}
\usepackage{xurl}
\usepackage{hyperref}

\newcolumntype{P}[1]{>{\raggedright\arraybackslash\footnotesize}m{#1}}
\newcolumntype{A}[1]{>{\centering\arraybackslash\footnotesize}m{#1}}

\usepackage[table,usenames,dvipsnames]{xcolor}
\definecolor{aa}{RGB}{175,238,238}
\definecolor{bb}{RGB}{255,255,255}

\usepackage{bm}
\usepackage{makecell}

\begin{document}
\title{SemVideoSte: Agentic AI for Cross-Modal Semantic Video Steganography Communication}

\author{Rui Meng,~\IEEEmembership{Member,~IEEE,} Song Gao, Tingting Zhu, Zixun Tang,
Jinke Ren,~\IEEEmembership{Member,~IEEE,}

Xiaodong Xu,~\IEEEmembership{Senior Member,~IEEE,} 
Pei Xiao,~\IEEEmembership{Senior Member,~IEEE,} 
Shuguang Cui,~\IEEEmembership{Fellow,~IEEE,}

and 
Ping Zhang,~\IEEEmembership{Fellow,~IEEE}

\thanks{

R. Meng, S. Gao, Z. Tang, X. Xu, and P. Zhang are with State Key Laboratory of Networking and Switching Technology, Beijing University of Posts and Telecommunications, Beijing 100876, China (e-mail: buptmengrui@bupt.edu.cn; wkd251292@bupt.edu.cn; 2023213283@bupt.cn; xuxiaodong@bupt.edu.cn; pzhang@bupt.edu.cn).
R. Meng and X. Xu are also with the Satellite Internet Testing Center, Xiong'an Aerospace Information Research Institute, Xiong'an 070001, China.

T. Zhu is with the ZGC Institute of Ubiquitous-X Innovation and Applications, Beijing 100083, China (e-mail: zhutingting@zgc-xnet.com).

J. Ren and S. Cui are with the School of Science and Engineering, the Shenzhen Future Network of Intelligence Institute, and the Guangdong Provincial Key Laboratory of Future Networks of Intelligence, The Chinese University of Hong Kong (Shenzhen), Shenzhen, Guangdong 518172, China (e-mail: jinkeren@cuhk.edu.cn; shuguangcui@cuhk.edu.cn).

P. Xiao is with Institute for Communication Systems (ICS), Home for 5GIC \& 6GIC, University of Surrey, Guildford, Surrey, GU2 7XH, United Kingdom (email: p.xiao@surrey.ac.uk).

}}

\maketitle

\begin{abstract}

Semantic Communication (SemCom) provides an efficient solution for supporting continuous video services in future \textit{Intellicise (intelligent and concise)} networks. 
In addition to public video content, such services often involve the exchange of sensitive auxiliary information such as identity credentials, target indices, and control instructions, while an additional encrypted transmission flow may expose the existence of confidential communication.
Semantic steganography provides a promising way to conceal such information within transmitted videos. Nevertheless, existing schemes mainly focus on text and image modalities, and their cover-edited embedding mechanisms cannot be directly extended to videos due to temporal inconsistency and semantic transmission distortion. 
To address these challenges, we propose an agentic artificial intelligence (AI)-driven cross-modal semantic video steganography communication (SemVideoSte) scheme for secure cross-modal information transmission. 
Agentic AI interprets the service context to determine a semantically matched public prompt and coordinates video generation, message embedding, semantic transmission, and message recovery process. A multi-mask mechanism combines variational autoencoder (VAE) reconstruction robustness, semantic transmission robustness, and latent separability to select reliable embedding positions. Meanwhile, a private digital token controls the random seed and guidance offset to construct positive and negative terminal latent states, enabling coverless message embedding during the final diffusion sampling step. 
Simulation results on open-source datasets demonstrate that SemVideoSte achieves high embedding capacity, reliable message recovery, and strong semantic consistency under different channel conditions. In addition, eavesdroppers without the correct digital token cannot reliably recover the secret message, and the generated stego videos effectively resist the steganalysis detection. We further establish a prototype testbed to verify the practical feasibility of SemVideoSte. The code is available at \url{https://github.com/kkir25/SemVideoSte-Agentic-AI-for-Cross-Modal-Semantic-Video-Steganography-Communication}.

\end{abstract}


\section{Introduction}

Future sixth-generation (6G) mobile communication networks are expected to support increasingly complex applications \cite{meng2026intellicise}, such as unmanned aerial systems, embodied agents, and immersive interaction \cite{zhang2024experts,11535035,zhang2024space-air-ground}, where numerous intelligent entities continuously perceive their surroundings, exchange information, and make decisions according to dynamically changing environments and service requirements \cite{zhang2026towards}. Compared with text and images, video simultaneously conveys scene context, object characteristics, motion states, and temporal evolution, making it an important modality for environmental perception and information interaction in 6G networks \cite{gao2025adaptive}.  However, the rapid growth in the number of connected devices and video resolutions leads to massive communication traffic and substantial transmission overhead.
\textit{Intellicise (intelligent and concise)} networks provide a promising paradigm for supporting such complex video transmission services \cite{meng2025semantic}. Different from conventional communication systems that mainly pursue accurate transmission of bit-level symbols, intellicise networks emphasize endogenous intelligence, semantic-aware information interaction, and native simplicity. Semantic Communication (SemCom), as a key enabling technology for intellicise networks, extracts and transmits task-relevant semantic representations instead of complete source video data, thereby reducing redundant information and improving transmission efficiency over bandwidth-limited and noisy wireless channels \cite{tang2024retrieval}. 
Therefore, SemCom provides an efficient solution for supporting continuous video services.

In addition to transmitting the video itself, intelligent entities often need to exchange auxiliary information during video services\cite{garcia2019essential}, such as identity credentials, device states, and target indices \cite{ko2021drone,zeng2019accessing,zhang2024interactive}. Although such information occupies substantially fewer bits than video data, it may contain sensitive task or confidential system information. For example, in surveillance and other mission-critical scenarios, an unmanned aerial platform may need to transmit confidential target indices and control instructions together with the video stream\cite{bennett2013analysis}, while cooperating robots may exchange sensitive identities and coordination commands\cite{zhang2025toward}. Directly transmitting these messages through an additional data stream not only consumes extra communication resources, but may also expose sensitive communication activities through traffic patterns and packet characteristics\cite{meng2025survey}. 

%
Conventional encryption methods protecs the content of confidential messages, but do not necessarily conceal their existence \cite{gao2026semstediff}. To this end, steganography embeds secret information into an inconspicuous carrier, allowing legitimate users to transmit confidential payloads without establishing an explicit information flow \cite{meng2025image}. When integrated with semantic video communication, the secret message could be hidden in the public video.
Even if an eavesdropper intercepts the transmitted semantic representation and reconstructs the visible video, the existence and content of the secret message still remain difficult to identify. Thus, semantic video steganography provides a suitable mechanism for jointly carrying public video semantics and confidential auxiliary information without arousing suspicions of potential eavesdroppers.


Existing semantic steganography communication schemes mainly focus on text and image modalities \cite{ni2025controllable,huo2025image,chencoding}, which provide limited space for covert message embeddings. In contrast, video steganography enables richer temporal characteristics to allow hidden information to be distributed across multiple frames\cite{kunhoth2023video}. 
Additionally, most image-oriented schemes follow a cover-edited paradigm, where confidential semantic information is embedded into pre-existing carriers through deep learning-based approaches, such as invertible neural networks (INNs)\cite{tang2025towards}. However, such methods are constrained by the quantity and resolution of the given carrier. More importantly, when extended to video communication, frame-wise modification may disturb temporal consistency and introduce abnormal variations in object appearance \cite{mou2023large}.
Motivated by coverless semantic steganography \cite{meng2026secure}, generative models provide a promising way to remove the dependence on pre-existing carriers.
In particular, generative diffusion models make it possible to synthesize the visible stego content and embed confidential information within the same generation process \cite{fan2026generative}.
Nevertheless, unlike image carriers with relatively fixed spatial structures, a generated video contains scene semantics,  motion evolution, and temporal consistency, which jointly determine whether the video is suitable for hiding information. 
To this end, the semantic video steganography should not only consider generating a visually natural stego video, but also ensure that the latent representation used for hiding information remains robust after semantic transmission.


Against this background, we introduce agentic artificial intelligence (AI) \cite{gao2025agentic} to autonomously achieve a cross-modal semantic video steganography. With the capability of interpreting service contexts, reasoning over task requirements, and invoking functional modules adaptively \cite{ jiang2025large}, agentic AI can connect video generation, hidden message embedding, and semantic transmission into a goal-oriented process. Rather than simply applying a fixed hiding rule to all generated videos, agentic AI is expected to coordinate the generation and embedding process according to the public service context, video generation characteristics, and transmission robustness requirements. In this way, the generated stego video can remain consistent with the target communication scenario, while the hidden information can be embedded into latent regions that are more suitable for reliable recovery after semantic transmission. 
The main contributions of this paper are summarized as follows:

\begin{itemize}
\item We propose an Agentic AI-driven cross-modal semantic video steganography communication (SemVideoSte) scheme, which conceals confidential textual auxiliary information within generated videos for secure transmission in video services. To the best of our knowledge, this is the first work to introduce semantic steganography into semantic video communication.
\item Motivated by \cite{wang2026ld}, we design a multi-mask-based latent position selection mechanism for reliable and secure secret message embedding. Specifically, a variational auto encoder (VAE)-based robust mask removes positions sensitive to latent-domain reconstruction, the semantic-based robust mask filters positions vulnerable to semantic codec and wireless channel distortion, and the separation mask preserves positions with distinguishable responses under token-controlled guidance. By combining these masks, SemVideoSte selects latent positions that are robust to semantic transmission and separable for message recovery.
\item We develop a classifier-free guidance-controlled message embedding mechanism based on the diffusion video generation process. The private digital token controls the random seed and guidance offset, thereby producing positive and negative terminal latent states for representing different secret bits. In this way, the secret message is embedded during the generation process rather than being modified by an existing video carrier, which enables natural and coverless stego video generation while reducing the risk of perceptible artifacts.
\item Simulation results on 100 public key prompts demonstrate the effectiveness and security of the proposed SemVideoSte. It achieves an average embedding capacity over 2200 bits, recovery accuracy of 99.62\%, and BERTScore F1 of 0.98, while maintaining a peak signal-to-noise ratio (PSNR) of 34.89~dB. Moreover, an agentic-AI based eavesdropper without the correct digital token achieves a recovery accuracy close to 50\%,  while the generated stego videos perform strong resistance to steganalysis.
\item We further develop a prototype validation platform, termed the SemVideoSte testbed, to evaluate the proposed scheme in a practical end-to-end deployment environment. Dedicated transmitter and receiver software interfaces are implemented to visualize model configurations, transmission statistics, runtime latency, secret-message recovery, and video reconstruction quality. The testbed provides both qualitative demonstrations and quantitative measurements, thereby validating the practical feasibility and effectiveness of the proposed SemVideoSte scheme.
\end{itemize}

\begin{figure*}
\centering
\includegraphics[width=\linewidth]{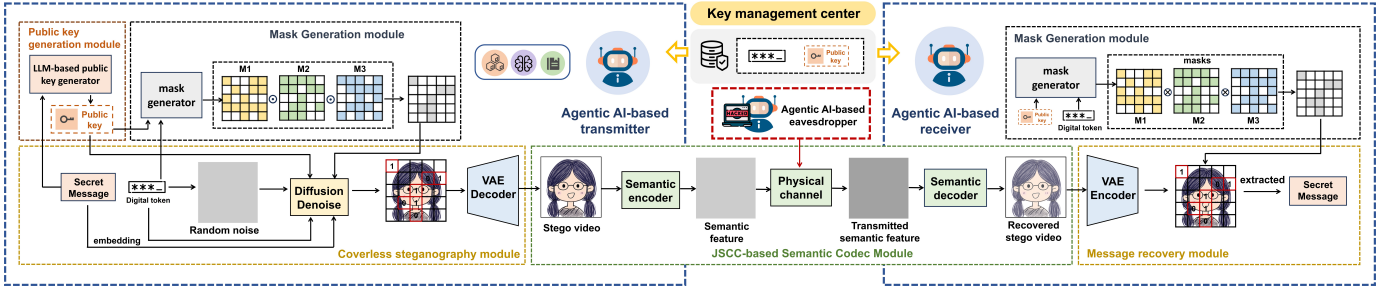}
\caption{The framework of proposed SemVideoSte scheme. The agentic AI-based transmitter generates a stego video under the control of the public key $K_{\rm pub}$ and private digital token $k$, and transmits its semantic representation through the wireless channel. The legitimate receiver reconstructs the stego video and recovers the secret message by regenerating the token-controlled embedding mask, while the eavesdropper attempts to infer the hidden message without the correct digital token.}
\label{fig_scheme}
\end{figure*}

\section{System Model}

\subsection{Network Model}
As illustrated in Figure \ref{fig_scheme}, we propose SemVideoSte to achieve video steganography, where the involved nodes are described as follows.

\subsubsection{Agentic AI-based Transmitter}

The agentic-AI based transmitter depends on the digital token $k$ to generate the embedding mask $\mathbf{M}_{\rm final}$. Then, it uses a public key $K_{\rm pub}$ and the digital token $k$ to guide the video diffusion process, while hiding secret message $m$ into the last sampling step with the control of embedding mask. After generating the stego video $\mathbf{x}_{\rm stego}$, the agentic-AI based transmitter encodes the video into semantic features for wireless transmission.

\subsubsection{Physical Channel}
The physical channel is modeled as aquasi-static fading channel with additive white Gaussian noise (AWGN) \cite{xu2023latent}. The stego video $\mathbf{x}_{stego}$ is transmitted through the channel as
\begin{equation}
\mathbf{x}’_{\rm stego} = h * \mathbf{x}_{\rm stego} + n,
\end{equation}
where $h$ is the coefficient of the physical channel, and $n$ is the noise component following a complex Gaussian distribution $\mathcal{CN}(0, \sigma_w^2)$ with $\sigma_w^2$ indicating the noise power. 

\subsubsection{Agentic AI-based Receiver}

Agentic AI-based receiver first reconstructs the transmitted stego video $\hat{\mathbf{x}}_{\rm stego}$ from the received semantic feature using the semantic decoder. After that, agentic AI maps $\hat{\mathbf{x}}_{\rm stego}$ back into the latent space and regenerates the same mask $\mathbf{M}_{\rm final}$ using the shared $K_{\rm pub}$ and digital token $k$. Based on the references of embedding mask, the receiver compares the values at specific positions and recovers the secret message as $\hat{\mathbf{m}}$. 

\subsubsection{Key Management Center}

The key management center stores and distributes the information required by legitimate agentic AI nodes. In the SemVideoSte framework, it is responsible for transmitting the public key $K_{\rm pub}$ and digital token $k$, which is assumed to provide a secure and reliable distribution channel for these two parameters. Significantly, the digital token $k$ is set to private. Even if the public key is exposed, the hidden message cannot be correctly extracted without the digital token $k$. 

\subsubsection{Agentic AI-based Eavesdropper}

The agentic AI-based eavesdropper intercepts the transmitted semantic representation and attempts to recover the secret message, which has access to the public generation model, semantic decoder module and even the public key $K_{\rm pub}$. A stronger eavesdropper may even infer the general workflow of the proposed SemVideoSte scheme. 


\subsection{Overview of the Proposed SemVideoSte Scheme}
As presented in Figure~\ref{fig_scheme}, the proposed SemVideoSte scheme consists of the following modules.

\subsubsection{Public Key Generation Module}

The public key generation module is used to generate a semantic prompt $K_{\rm pub}$ to determine the content of stego video. Based on the perceived transmission message $m$ and transmission requirement $\mathcal{R}$, agentic AI invokes the LLM-based public key generator $f_{\rm key}(\cdot)$ to produce and select a suitable prompt, which is responsible for generating the subsequent stego video with stable visual content and motion pattern. 
This process can be expressed as
\begin{equation}
    K_{\rm pub} = f_{\rm key}(m,\mathcal{R} ).
\end{equation}

\subsubsection{Mask Generation Module}
The mask generation module is responsible for selecting reliable positions for the secret message embedding. Agentic AI generates three different masks at both transmitter and receiver with shared public key and digital token, where the VAE-based robust mask $\mathbf{M}_{\rm vae}$ removes positions that are unstable during reconstruction from bit-domain to latent domain, the semantic-based robust mask $\mathbf{M}_{\rm sem}$ removes positions that are sensitive to semantic encoding and decoding process, and the separation mask $\mathbf{M}_{\rm sep}$ preserves positions that have distinguishable latent responses under conditional sampling process. By combining these constraints, the final embedding mask $ \mathbf{M}_{\rm final}$ is expressed as
\begin{equation}
\label{final_mask}
    \mathbf{M}_{\rm final}
    =
    \mathbf{M}_{\rm vae}
    \odot
    \mathbf{M}_{\rm sem}
    \odot
    \mathbf{M}_{\rm sep},
\end{equation}
where $\odot$ denotes element-wise multiplication. Depending on $\mathbf{M}_{\rm final}$, agentic AI can only embed secret message into positions that are robust to transmission distortion and separable for message recovery.

\subsubsection{Cross-Modal Coverless Steganography Module}
To realize invisible message embedding, agentic AI relies on conditional diffusion module for coverless stego video generation. At the transmitter, the digital token $k$ initializes the random noise and controls the conditional sampling trajectory. Meanwhile, the public key $K_{\rm pub}$ is applied as the condition to guide the generation of the video content. During the final sampling stage, the secret message $\mathbf{m}$ is embedded into the selected latent positions under the control of $\mathbf{M}_{\rm final}$. The stego latent representation $\mathbf{z}_{\rm stego}$ can be generally expressed as
\begin{equation}
    \mathbf{z}_{\rm stego}
    =
    \mathcal{F}
    \left(
    K_{\rm pub}, k, \mathbf{m}, \mathbf{M}_{\rm final}
    \right).
\end{equation}
Then, the VAE decoder $\mathcal{D}_{\rm VAE}(\cdot)$ converts the stego latent vector $\mathbf{z}_{\rm stego}$ into the stego video $\mathbf{x}_{\rm stego}$ by
\begin{equation}
    \mathbf{x}_{\rm stego}
    =
    \mathcal{D}_{\rm VAE}
    \left(
    \mathbf{z}_{\rm stego}
    \right).
\end{equation}

\subsubsection{JSCC-based Semantic Codec Module}

The JSCC-based semantic codec module is employed to transmit the generated stego video through the wireless channel, which is considered as a typical SemCom transmission. At the transmitter side , the semantic encoder $\mathcal{E}_{\rm sem}(\cdot)$ maps the stego video $\mathbf{x}_{\rm stego}$ into the semantic feature $\mathbf{s}_{\rm stego}$ by
\begin{equation}
    \mathbf{s}_{\rm stego}
    =
    \mathcal{E}_{\rm sem}
    \left(
    \mathbf{x}_{\rm stego}
    \right).
\end{equation}
After passing through the physical channel, the received stego video $\hat{\mathbf{x}}_{\rm stego}$ is recovered from the noisy semantic feature $\hat{\mathbf{s}}_{\rm stego}$ by the semantic decoder $\mathcal{D}_{\rm sem}(\cdot)$ employed at the receiver side as
\begin{equation}
    \hat{\mathbf{x}}_{\rm stego}
    =
    \mathcal{D}_{\rm sem}
    \left(
    \hat{\mathbf{s}}_{\rm stego}
    \right).
\end{equation}

\subsubsection{Message Recovery Module}

The message recovery module is deployed at the receiver side to extract the hidden message from the recovered stego video. After semantic video transmission, the agentic AI-based receiver invokes the VAE encoder $\mathcal{E}_{\rm VAE}(\cdot)$ to transform the recovered stego video $\hat{\mathbf{x}}_{\rm stego}$ into the latent domain as
\begin{equation}
    \hat{\mathbf{z}}_{\rm stego} = \mathcal{E}_{\rm VAE}(\hat{\mathbf{x}}_{\rm stego}).
\end{equation}
Based on the shared public key $K_{\rm pub}$ and the private digital token $k$, agentic AI regenerates the final embedding mask $\mathbf{M}_{\rm final}$ and the corresponding token-controlled reference states through the mask generation module, which provide the decision reference for distinguishing different embedded bits at the masked positions. The mask-based recovery process can be shown as
\begin{equation}
    \hat{\mathbf{m}} =
    {f}_{\rm msg}
    \left(
    \hat{\mathbf{z}}_{\rm stego},
    \mathbf{z}_{T}^{+},
    \mathbf{z}_{T}^{-},
    \mathbf{M}_{\rm final}
    \right),
\end{equation}
where ${f}_{\rm msg}(\cdot)$ denotes the message recovery operator, and $\mathbf{z}_{T}^{+}$ and $\mathbf{z}_{T}^{-}$ are the positive and negative latent references, respectively. 

\subsection{Workflow of Agentic AI}
\begin{figure*}[htbp]
\centering
\includegraphics[width=0.9\linewidth]{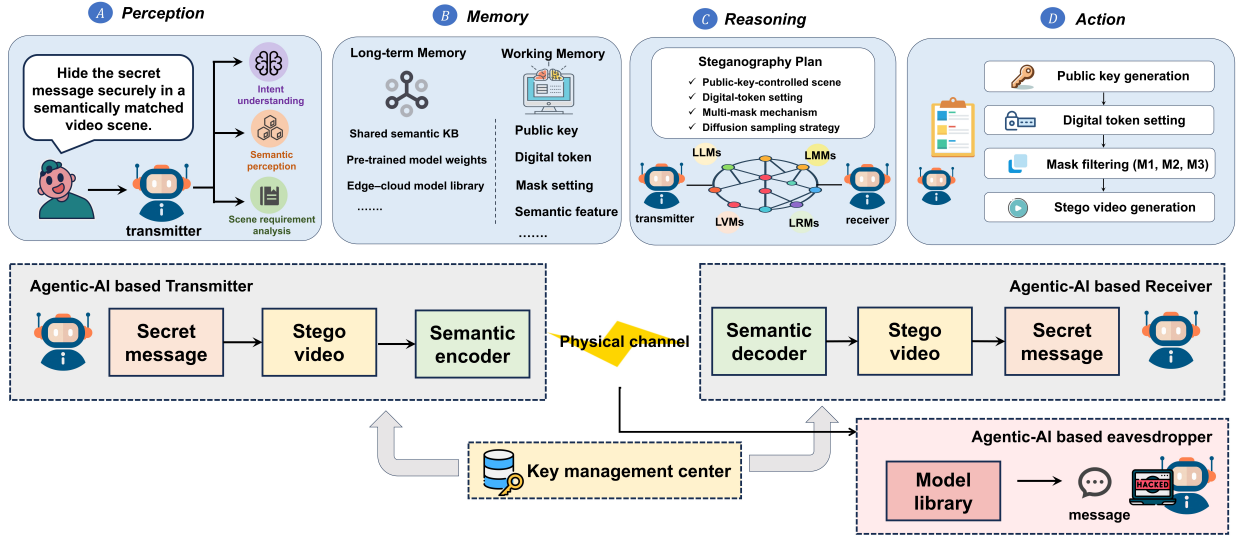}
\caption{Agentic AI workflow in the proposed SemVideoSte framework. Agentic AI performs perception, memory , reasoning, and action to coordinate public key generation, digital token setting, mask filtering, stego video generation, semantic transmission, and secret message recovery to achieve semantic video
steganography communication.}
\label{fig_framework}
\end{figure*}
As shown in Figure~\ref{fig_framework}, agentic AI follows a four-stage workflow, including perception, memory, reasoning, and action \cite{abou2025agentic}. Rather than executing a fixed steganography pipeline, it behaves as an autonomous coordinator that interprets the hiding intention, retrieves task-related knowledge, plans the steganography strategy, and invokes functional modules to complete secure video generation and message recovery.

\begin{itemize}
\item \textit{Step 1: Perception for Intent Understanding and Scenario Analysis:}
Upon receiving the objective of hiding the secret message securely in a video scene, agentic AI first interprets the transmission intention and task constraints. This stage includes understanding the secret message requirement, identifying the target communication scenario, and analyzing the expected visual semantics of the generated video. Through semantic perception and scene requirement analysis, agentic AI abstracts the task into a secure steganography objective, where the generated stego video should be natural, context-consistent, and inconspicuous to potential eavesdroppers. In this design, the framework leverages a public key to control the semantic content and motion of the generated video, and uses a private digital token to guide the sampling trajectory and embedding positions for secure message hiding.

\item \textit{Step 2: Memory Retrieval and Task State Organization:}
After perception, agentic AI retrieves useful knowledge from both long-term memory and working memory\cite{zhang2026toward}. Long-term memory contains the shared semantic knowledge base, pre-trained model weights, edge-cloud model library, and historical experience of video generation and steganography. Working memory records the temporary task states of the current transmission, including the public key, digital token, mask setting, semantic feature, and channel-related information. This memory mechanism enables agentic AI to connect the current hiding task with prior knowledge, avoiding isolated decisions based only on the input message.

\item \textit{Step 3: Reasoning for Steganography Strategy Planning:}
Based on the perceived task and retrieved memory, agentic AI performs reasoning to formulate a cross-modal steganography plan. Specifically, it determines what kind of public-key-controlled scene is suitable for the current transmission scenario, how the digital token should control the sampling trajectory, and how the multi-mask mechanism should balance robustness and separability. During this process, agentic AI jointly considers covertness, embedding capacity, transmission robustness, and recovery reliability, so that the generated video can hide the message without causing abnormal semantic or visual patterns.

\item \textit{Step 4: Action Through Module Invocation and Feedback:}
Finally, agentic AI executes the planned strategy by invoking different modules in sequence. At the transmitter, it triggers public key generation, digital token setting, mask filtering, and stego video generation. The generated stego video is then transmitted through the semantic codec and physical channel. At the receiver, agentic AI invokes the corresponding recovery process, including semantic decoding, mask regeneration, and masked-position message extraction. The execution results can be further fed back to the working memory, supporting future adjustment of public key selection, digital token setting, and mask filtering.
\end{itemize}


\section{Public key Generation Module}
In the proposed SemVideoSte scheme, the public key is defined as a textual prompt that controls the semantic content and motion pattern of the generated stego video. Different from conventional cryptographic keys\cite{tung2023deep}, the public key does not contain secret bits and is not used to directly encrypt the private message. Instead, it serves as a public semantic condition to determine the visible content and motion pattern of the stego video \cite{gao2026semstediff}. To generate a suitable public key, we introduce an agentic AI-driven public key generation module.

\subsection{Semantic-Matched Prompt Generation }

Since the stego video is transmitted over an open wireless environment, its visible content needs to be consistent with the target communication scenario, because an intelligent eavesdropper may identify the generated video as abnormal due to semantic inconsistency \cite{tang2026eavesdroppers}. Therefore, agentic AI should determine the public key based on the perceived transmission message. 

Let $\mathcal{C}$ denote the public context perceived by agentic AI, which includes the service scenario, environmental background, object category, and task requirement. Agentic AI invokes a LLM  to generate a candidate prompt set as
\begin{equation}
    \mathcal{P} = \{P_1, P_2, \ldots, P_N\}
    = f_{\rm LLM}(\mathcal{C}),
\end{equation}
where $P_i$ denotes the $i$-th candidate public key. 
For video generation, a public prompt should contain not only static scene semantics but also motion semantics. Thus, we have
\begin{equation}
    P_i = \{p_i^{\rm scene}, p_i^{\rm obj}, p_i^{\rm motion}\},
\end{equation}
where $p_i^{\rm scene}$ describes the background environment, $p_i^{\rm obj}$ describes the main objects, and $p_i^{\rm motion}$ describes the temporal evolution of the video.

\subsection{Carrier-Aware Prompt Selection}

Although the candidate prompts generated by LLMs are semantically matched with the transmission scenario, not all of them are suitable for video steganography. In our scheme, the public key not only controls the visible semantics of the stego video, but also determines the latent sampling trajectory of the video diffusion model. Therefore, an appropriate public key should generate a stego video with stable visual content and smooth temporal motion, which can achieve robust message embedding and reconstruction in SemCom \cite{tan2021coverless}.

To this end, agentic AI further performs carrier-aware prompt selection based on the candidate prompt set $\mathcal{P}$ \cite{meng2026secure}. For each $P_i$, agentic AI generates a conditional video and perceives its carrier characteristics such as visual naturalness, embedding capacity, and so on. This process is generally expressed as
\begin{equation}
    K_{\rm pub} = \mathcal{A}_{\rm select}\left(\mathcal{P} \mid \mathcal{Q}_{\rm car}\right),
\end{equation}
where $\mathcal{A}_{\rm select}(\cdot)$ denotes the prompt selection implemented by agentic AI, and $\mathcal{Q}_{\rm car}$ represents the carrier-aware requirements.
Specifically, prompts with chaotic object motion or unstable textures should be avoided, which may lead to fragile latent trajectories and unreliable bit recovery \cite{meng2023robust}. In contrast, prompts that produce temporally smooth videos with consistent objects and moderate motion are preferred, because they can provide more stable carrier positions for secret message embedding. The selected public key $K_{\rm pub}$ is used as the public condition for the following mask generation and semantic video steganography modules.

\begin{figure}
\centering
\includegraphics[width=\linewidth]{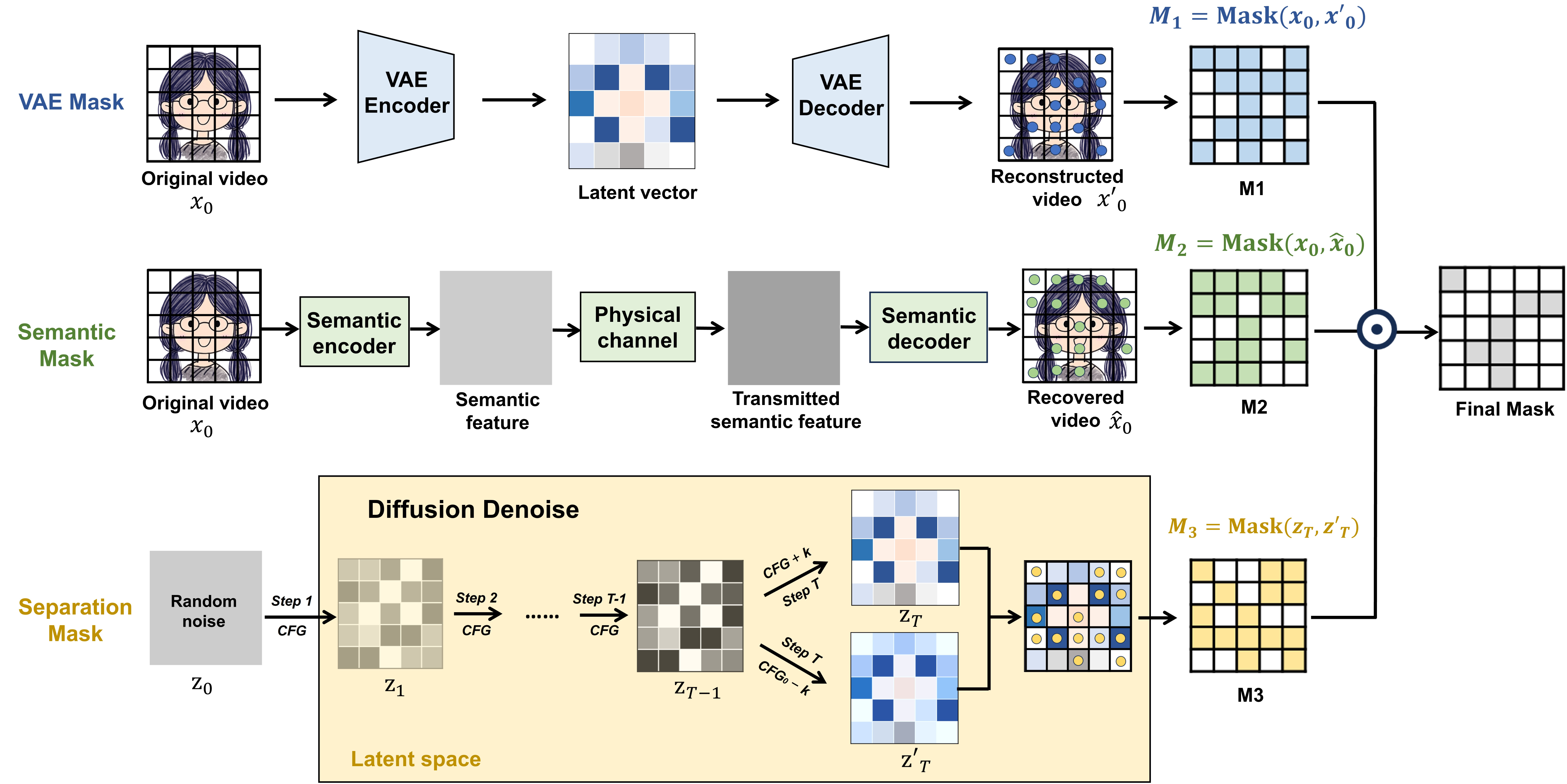}
\caption{The generation process of the final embedding mask. Three masks are constructed to evaluate VAE reconstruction robustness, semantic transmission robustness, and latent separation under token-controlled guidance, respectively. Their intersection forms the final mask $\mathbf{M}_{\rm final}$, which selects reliable and separable latent positions for secret message embedding.}
\label{fig_mask}
\vspace{-10pt}
\end{figure}

\section{Mask Generation Module}
As illustrated in Figure \ref{fig_mask}, the embedded message should remain robust during VAE reconstruction and semantic video transmission, while different message bits should also be distinguishable in the diffusion latent space. To achieve this objective, agentic AI constructs three masks, including the VAE-based robust mask, semantic-based robust mask, and separation mask. These masks are combined to obtain the available embedding positions.

\subsection{Robust Mask Generation}
The robust mask generation focuses on the stability of carrier positions under unavoidable reconstruction and transmission distortion. Agentic AI first perceives the generated video and then invokes the VAE module and semantic video transmission module to identify robust regions \cite{kingma2013auto}. The VAE-based robust mask evaluates the distortion caused by the VAE encoder-decoder process, while the semantic-based robust mask evaluates the distortion caused by semantic transmission over wireless channels.
\subsubsection{VAE-based Robust Mask}
In video diffusion models, the generated video is represented and processed in the latent space. The conversion between the pixel domain and latent domain may introduce local reconstruction errors, which can destroy the embedded message even before wireless transmission. Thus, agentic AI considers the reconstruction stability of the generated video through the VAE module.

First, agentic AI obtains the VAE-reconstructed vector of the original generative video $\mathbf{x}_0$ by
\begin{equation}
    \mathbf{x}'_0 = \mathcal{D}_{\rm VAE}\left(\mathcal{E}_{\rm VAE}(\mathbf{x}_0)\right),
\end{equation}
where $\mathcal{E}_{\rm VAE}(\cdot)$ and $\mathcal{D}_{\rm VAE}(\cdot)$ denote the VAE encoder and decoder, respectively. The VAE reconstruction difference $\mathbf{E}_{\rm vae}$ between $\mathbf{x}_0$ and $\mathbf{x}'_0$ is measured as
\begin{equation}
    \mathbf{E}_{\rm vae} = \left|\mathbf{x}_0 - \mathbf{x}'_0\right|.
\end{equation}
Based on the perceived reconstruction error, agentic AI generates the VAE-based robust mask as
\begin{equation}
    \mathbf{M}_{\rm vae} = \mathbb{I}\left(\mathbf{E}_{\rm vae} \leq \tau_{\rm vae}\right),
\end{equation}
where $\mathbb{I}(\cdot)$ denotes the indicator function and $\tau_{\rm vae}$ is the VAE robustness threshold. The positions retained by $\mathbf{M}_{\rm vae}$ are less affected by VAE reconstruction, which are more suitable for message embedding.

\subsubsection{Semantic-based Robust Mask}

The semantic-based robust mask is designed to select positions that remain stable through semantic video transmission. In SemVideoSte, the stego video is transmitted through the semantic encoder, physical channel, and semantic decoder. Channel noise and semantic reconstruction errors may change local visual details, causing bit recovery errors at the receiver \cite{meng2025secure}. To address such distortion, agentic AI evaluates the transmission stability of the generated video before message embedding.
The generated video $\mathbf{x}_0$ is encoded into semantic features and transmitted through the physical channel. The reconstructed video is obtained by
\begin{equation}
    \hat{\mathbf{x}}_0 =
    \mathcal{D}_{\rm sem}
    \left(
    \mathcal{H}
    \left(
    \mathcal{E}_{\rm sem}(\mathbf{x}_0)
    \right)
    \right).
\end{equation}
Then, agentic AI measures the semantic transmission distortion $\mathbf{E}_{\rm sem}$ and generates semantic-based robust mask as 
\begin{equation}
\label{Esem}
    \mathbf{E}_{\rm sem} = \left|\mathbf{x}_0 - \hat{\mathbf{x}}_0\right|,
\end{equation}

\begin{equation}
\label{Msem}
    \mathbf{M}_{\rm sem} = \mathbb{I}\left(\mathbf{E}_{\rm sem} \leq \tau_{\rm sem}\right),
\end{equation}
where $\tau_{\rm sem}$ denotes the semantic robustness threshold. 

\begin{rem}
Through this mask, the retained positions are more stable after semantic video transmission, which improves the reliability of message recovery under noisy wireless channels.
\end{rem}
\subsection{Separation Mask Generation}
While the robust masks can reduce bit errors caused by wireless transmission, they do not guarantee that different secret bits can be correctly distinguished \cite{wang2026ld}. As the message embedding is achieved by controlling the diffusion sampling trajectory through classifier-free guidance, agentic AI further designs a separation mask to select positions that show clear differences under opposite guidance.
During diffusion denoising, agentic AI performs two conditional sampling processes under the same public key. The digital token $k$ controls the guidance response of the sampling trajectory, which leads to the positive and negative guidance states as
\begin{equation}
    cfg_k^{+} = cfg + \Delta cfg_k,
\end{equation}
\begin{equation}
    cfg_k^{-} = cfg - \Delta cfg_k,
\end{equation}
where $cfg$ is the original classifier-free guidance scale\cite{ho2022classifier}, and $\Delta cfg_k$ is the token-controlled guidance offset. The two conditioned sampling states are further defined as
\begin{equation}
    \mathbf{c}_{k}^{+} = \left(k, cfg_k^{+}\right),
\end{equation}
\begin{equation}
    \mathbf{c}_{k}^{-} = \left(k, cfg_k^{-}\right).
\end{equation}
Then, agentic AI generates these two reversed sampling latent states $\mathbf{c}_{k}^{+}$ and $\mathbf{c}_{k}^{-}$  as

\begin{equation}
    \mathbf{z}_{T}^{+} = \mathcal{F}(K_{\rm pub}, \mathbf{c}_{k}^{+}) = \mathcal{F}(K_{\rm pub}, k, cfg_k^{+}),
\end{equation}
\begin{equation}
    \mathbf{z}_{T}^{-} = \mathcal{F}(K_{\rm pub}, \mathbf{c}_{k}^{-}) = \mathcal{F}(K_{\rm pub}, k,cfg_k^{-}).
\end{equation}
Under this token-controlled conditional sampling mechanism, the two latent states share the same public semantic condition but exhibit opposite trajectory responses. The positions with larger response differences are more suitable for binary message embedding, because they provide stronger distinguishability during message recovery. The latent separation is calculated by
\begin{equation}
    \mathbf{E}_{\rm sep} = \left|\mathbf{z}_{T}^{+} - \mathbf{z}_{T}^{-}\right|.
\end{equation}

Then, agentic AI generates the separation mask as
\begin{equation}
    \mathbf{M}_{\rm sep} = \mathbb{I}\left(\mathbf{E}_{\rm sep} \geq \tau_{\rm sep}\right),
\end{equation}
where $\tau_{\rm sep}$ denotes the separation threshold. 
\begin{rem}
After the three masks are obtained, agentic AI combines them to determine the final embedding mask by (\ref{final_mask}). The final mask only preserves positions that are robust to VAE reconstruction and semantic transmission, while also separable under different guidance, which ensures the successful embedding of secret message.
\end{rem}
\section{Cross-Modal Semantic Video Steganography Module}
The cross-modal semantic video steganography module is employed to realize the embedding and recovery of secret message from the semantic video, which is divided into conditional diffusion model-based coverless video generation at the transmitter and mask-based secret message recovery at the receiver.
\subsection{Conditional Diffusion Model-based Coverless Video Generation}
\begin{figure}
\centering
\includegraphics[width=\linewidth]{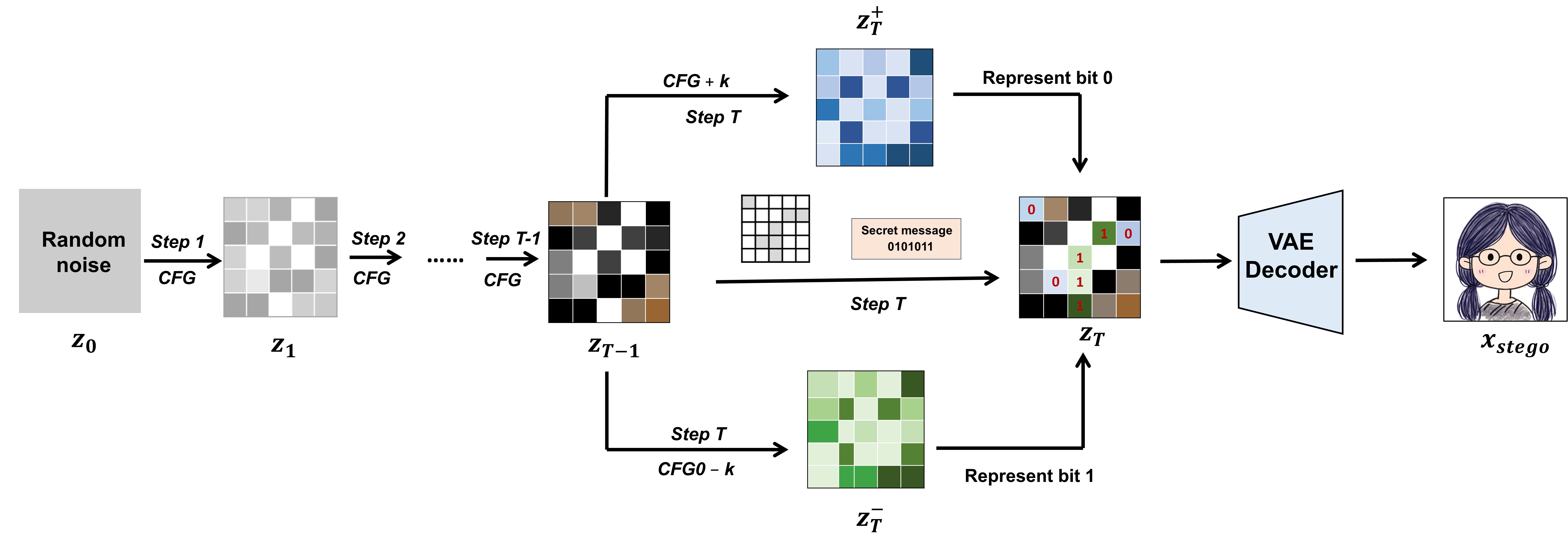}
\caption{The process of classifier-free guidance controlled message embedding. The positive and negative token-controlled guidance states generate two separable latent states, which are used to represent different secret bit values.}
\label{fig_embed}
\vspace{-10pt}
\end{figure}

As shown in Figure \ref{fig_embed}, the transmitter considers two steps to obtain the visually natural stego video embedded with secret message, including flow matching-based sampling and classifier-free guidance controlled message embedding.

\subsubsection{Flow Matching-based Sampling}

To generate stego videos for message hiding, agentic AI adopts a conditional diffusion process. Compared with image steganography, video steganography requires not only frame-level visual naturalness but also inter-frame motion consistency. If the latent generation process is unstable, the generated video may suffer from visual blurring, object distortion, or irregular motion, which can arouse suspicion of eavesdroppers and impact accuracy of bit recovery. Therefore, we employ flow matching for diffusion sampling. Different from conventional diffusion models that predict the noise component at each denoising step, it directly predicts the direction of the generative trajectory, which provides a smooth and controllable generative path for coverless video steganography \cite{dao2023flow}.

Agentic AI first samples the initial latent noise under the control of the digital token. Specifically, the digital token $k$ is mapped into a deterministic random seed by $s = \mathcal{\text{Hash}}(k)$. Agentic AI samples the initial video latent noise as
\begin{equation}
    \mathbf{z}_{0} = \mathrm{Randn}(s),
\end{equation}
where $\mathrm{Randn}(s)$ denotes Gaussian noise sampling under seed $s$. Since the same digital token leads to the same initial latent noise, the legitimate transmitter and receiver can generate the same starting point of the video generation trajectory.

After obtaining the initial noise, agentic AI performs flow matching-based sampling to gradually transform the noisy latent state into a video latent representation \cite{rombach2022high}. Under the condition of $K_{\rm pub}$, the video diffusion transformer predicts a velocity field to guide the latent state $\mathbf{z}_{t}$ from $\mathbf{z}_{0}$ to the target video latent distribution by
\begin{equation}
    \mathbf{v}_{t}
    =
    \frac{d\mathbf{z}_{t}}{dt}
    =
    \mathbf{v}_{\theta}
    \left(
    \mathbf{z}_{t}, t, K_{\rm pub}
    \right),
\end{equation}
where $\mathbf{v}_{\theta}(\cdot)$ is the velocity predicted by the video diffusion transformer.

Then, agentic AI determines the total sampling steps $T$ and the sampling shift factor $\mu$ to discretize the continuous flow trajectory. The agentic AI first constructs a uniform noise-level sequence $\sigma_t$ by
\begin{equation}
    \sigma_t = 1-\frac{t}{T},
    \quad t=0,1,\ldots,T.
\end{equation}
A larger $\sigma_t$ indicates the latent state is closer to the initial noisy distribution, while a smaller $\sigma_t$ indicates its closer to clean video latent representation. Meanwhile, the sampling shift factor $\mu$ is further used to adjust how these sampling points are distributed along the noise trajectory, which aims to balance global motion consistency and local visual fidelity during video generation \cite{lipman2022flow}. Thus, we have
\begin{equation}
    \sigma_t'
    =
    \frac{\mu \sigma_t}
    {1+(\mu-1)\sigma_t},
\end{equation}
where $\sigma_t'$ is the shifted noise level.

The latent state is updated along the flow trajectory by
\begin{equation}
\label{flow_upodate}
    \mathbf{z}_{t+1}
    =
    \mathbf{z}_{t}
    +
    \Delta \sigma_t'
    \mathbf{v}_{t},
\end{equation}
where $\Delta \sigma_t' = \sigma_{t+1}'-\sigma_t'$ is the interval between two shifted noise levels.

Based on this flow matching-based sampling process, agentic AI obtains a natural video latent vector with appropriate semantic content and stable temporal motion, which provides the latent foundation for the following guidance-controlled message embedding.

\subsubsection{Classifier-Free Guidance Controlled Message Embedding}
Agentic AI further embeds secret bits by controlling the classifier-free guidance response at the final sampling stage \cite{ho2022classifier}. 
At the $t$-th step, the video diffusion transformer predicts the conditional velocity and unconditional velocity as

\begin{equation}
    \mathbf{v}_{t}^{\rm cond}
    =
    \mathbf{v}_{\theta}
    \left(
    \mathbf{z}_{t}, t, K_{\rm pub}
    \right),
\end{equation}
\begin{equation}
    \mathbf{v}_{t}^{\rm uncond}
    =
    \mathbf{v}_{\theta}
    \left(
    \mathbf{z}_{t}, t, \varnothing
    \right),
\end{equation}
where $\varnothing$ denotes the empty condition. Then, the guided velocity is calculated by
\begin{equation}
    \mathbf{v}_{t}^{\rm cfg}
    =
    \mathbf{v}_{t}^{\rm uncond}
    +
    cfg
    \left(
    \mathbf{v}_{t}^{\rm cond}
    -
    \mathbf{v}_{t}^{\rm uncond}
    \right).
\end{equation}
The guidance term determines the semantic strength of the public key during diffusion sampling.

To preserve the global structure and motion consistency of the generated video, agentic AI first performs the standard flow matching sampling process from $t=0$ to $t=T-1$. At the final sampling step, the latent state $\mathbf{z}_{T-1}$ is used to construct two candidate terminal latent states. The positive and negative guidance velocity are calculated respectively as
\begin{equation}
    \mathbf{v}_{T-1}^{+}
    =
    \mathbf{v}_{T-1}^{\rm uncond}
    +
    cfg_k^{+}
    \left(
    \mathbf{v}_{T-1}^{\rm cond}
    -
    \mathbf{v}_{T-1}^{\rm uncond}
    \right),
\end{equation}
\begin{equation}
    \mathbf{v}_{T-1}^{-}
    =
    \mathbf{v}_{T-1}^{\rm uncond}
    +
    cfg_k^{-}
    \left(
    \mathbf{v}_{T-1}^{\rm cond}
    -
    \mathbf{v}_{T-1}^{\rm uncond}
    \right).
\end{equation}

According to (\ref{flow_upodate}), the corresponding positive and negative terminal latent states are expressed as
\begin{equation}
    \mathbf{z}_{T}^{+}
    =
    \mathbf{z}_{T-1}
    +
    \Delta \sigma_{T-1}'
    \mathbf{v}_{T-1}^{+},
\end{equation}
\begin{equation}
    \mathbf{z}_{T}^{-}
    =
    \mathbf{z}_{T-1}
    +
    \Delta \sigma_{T-1}'
    \mathbf{v}_{T-1}^{-}.
\end{equation}
While $\mathbf{z}_{T}^{+}$ and $\mathbf{z}_{T}^{-}$ are derived from the same previous latent state $\mathbf{z}_{T-1}$, their difference mainly comes from the token-controlled  guidance. Therefore, they can be regarded as two latent codewords to represent 1 and 0. 

By transforming secret message $\mathbf{m}$ into bit-domain, it can be represented as
\begin{equation}
    \mathbf{b}
    =
    \{b_1,b_2,\ldots,b_L\},
    \quad b_l \in \{0,1\}.
\end{equation}

Based on the final embedding mask $\mathbf{M}_{\rm final}$, agentic AI obtains the reliable embedding position set
\begin{equation}
    \Omega=\{p_1,p_2,\ldots,p_L\},
\end{equation}
where $p_l$ denotes the $l$-th selected latent position. For the secret bit $b_l$, agentic AI embeds it by selecting the corresponding latent value from the negative or positive terminal latent state at the final sampling step as
\begin{equation}
    \mathbf{z}_{\rm stego}(p_l)
    =
    \left(1-b_l\right)
    \mathbf{z}_{T}^{-}(p_l)
    +
    b_l
    \mathbf{z}_{T}^{+}(p_l).
\end{equation}
It can also be expressed as
\begin{equation}
    \mathbf{z}_{\rm stego}(p_l)
    =
    \begin{cases}
    \mathbf{z}_{T}^{-}(p_l), & b_l=0,\\
    \mathbf{z}_{T}^{+}(p_l), & b_l=1.
    \end{cases}
\end{equation}

For positions outside the selected embedding position set $\Omega$, the latent values remain unchanged from the normally generated terminal latent state. Thus, the embedding operation is restricted to the reliable positions determined by $\mathbf{M}_{\rm final}$, while the remaining latent components preserve the original video semantics and temporal structure.

Finally, agentic AI decodes the stego video latent vector $\mathbf{z}_{\rm stego}$ into the pixel domain by invoking the VAE decoder on the stego video latent state

\begin{equation}
    \mathbf{x}_{\rm stego}
    =
    \mathcal{D}_{\rm VAE}
    \left(
    \mathbf{z}_{\rm stego}
    \right),
\end{equation}
where $\mathcal{D}_{\rm VAE}(\cdot)$ denotes the VAE decoder. 

\begin{rem}
Through this classifier-free guidance controlled embedding process, the secret bits are implicitly represented by positive and negative terminal latent responses under the same public semantic condition. Since the selected positions are robust to reconstruction and transmission distortion and also separable under different guidance states, the proposed scheme can embed secret messages into generated videos with strong covertness and recovery reliability.
\end{rem}

\subsection{Mask-based Secret Message Recovery}
\begin{figure}
\centering
\includegraphics[width=\linewidth]{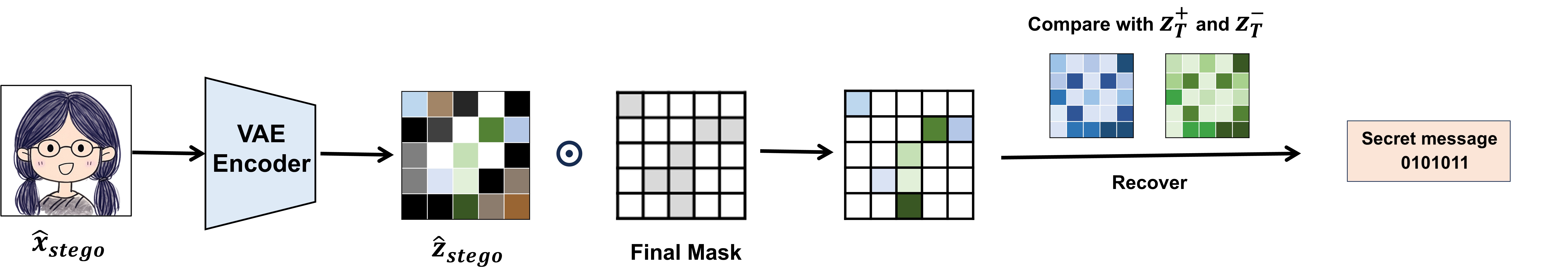}
\caption{The process of mask-based secret message recovery, where the received stego video is encoded into the latent space and the latent values at the masked positions are extracted. Then, the specific values are compared with the regenerated positive and negative reference states to determine the embedded bits and recover the secret message.}
\label{fig_recovery}
\vspace{-10pt}
\end{figure}

At the receiver, agentic AI recovers the secret message from the received stego video based on the same public key, digital token, and final embedding mask, which is shown in Figure \ref{fig_recovery}.  For the secret bits are embedded in the video latent space, agentic AI first maps the received stego video $\hat{\mathbf{x}}_{\rm stego}$ back into the latent domain by
\begin{equation}
    \hat{\mathbf{z}}_{\rm stego}
    =
    \mathcal{E}_{\rm VAE}
    \left(
    \hat{\mathbf{x}}_{\rm stego}
    \right),
\end{equation}
where $\mathcal{E}_{\rm VAE}(\cdot)$ denotes the VAE encoder.

Then, agentic AI regenerates the positive latent state $\mathbf{z}_{T}^{+}$ and negative one $\mathbf{z}_{T}^{-}$ using the same public key and digital token sharing by a key management center. Also, the final embedding mask $\mathbf{M}_{\rm final}$ is generated locally. For each selected position $p_l$, agentic AI calculates the distance between the received latent value and the two latent codewords as
\begin{equation}
    d_l^{-}
    =
    \left|
    \hat{\mathbf{z}}_{\rm stego}(p_l)
    -
    \mathbf{z}_{T}^{-}(p_l)
    \right|,
\end{equation}
\begin{equation}
    d_l^{+}
    =
    \left|
    \hat{\mathbf{z}}_{\rm stego}(p_l)
    -
    \mathbf{z}_{T}^{+}(p_l)
    \right|.
\end{equation}
The embedded bit is recovered according to the minimum-distance decision rule as
\begin{equation}
    \hat{b}_l
    =
    \begin{cases}
    0, & d_l^{-} \leq d_l^{+},\\
    1, & d_l^{+} < d_l^{-}.
    \end{cases}
\end{equation}
After all selected positions are decoded, the recovered bit sequence is obtained as
\begin{equation}
    \hat{\mathbf{b}}
    =
    \{\hat{b}_1,\hat{b}_2,\ldots,\hat{b}_L\}.
\end{equation}
The recovered bit sequence can be further converted into the recovered message $\hat{\mathbf{m}}$.

\begin{rem}
The above recovery process shows that the secret message can be extracted without estimating the generative trajectory from the received video. Agentic AI can only compare the received latent values with the regenerated positive and negative latent vectors at the masked positions in $\Omega$. This lightweight recovery mechanism reduces the decoding complexity and avoids dependence on full trajectory inversion. Moreover, by restricting the decision process to the masked robust and separable latent positions, the proposed scheme improves the recovery reliability under semantic transmission distortion while preserving the coverless nature of video generation.
\end{rem}

\section{Simulation Results and Analysis}
\subsection{Simulation Parameters} \label{args}
\subsubsection{Parameter Settings}
For the public key generation module, agentic AI invokes the ChatGPT-5.4 API to generate semantic-matched public prompts according to the secret message. For the coverless steganography module, Wan2.1-T2V-1.3B\footnote{\url{https://huggingface.co/Wan-AI/Wan2.1-T2V-1.3B}} is employed as the text-to-video diffusion generator. The generated video resolution is set to $832 \times 480$ with 81 frames. The sampling solver is UniPC, where the number of sampling steps is set to 50 and the sample shift is 5.0. The seed $s$ of diffusion process is set as 99. Also, the basic guidance scale of the video diffusion model is set to 5.0. For the JSCC-based semantic codec module, the proposed SemVideoSte scheme is verified under the SwinJSCC backbone \cite{yang2024swinjscc}, which is trained on the CelebA-HQ dataset \cite{karras2017progressive} under the AWGN channel fixed at $\text{SNR} = 10$ dB. The channel bandwidth parameter is set to $C=48$. It is noted that although the transmitted carrier is a video sequence, we adopt a single-frame semantic transmission strategy in the current simulation to separate channel distortion from variations introduced by temporal prediction. Specifically, each video frame is independently encoded and decoded by the image-based JSCC model, corresponding to the group-of-picture (GOP) setting as 1. The training of JSCC models and testing of SemSteDiff is performed on NVIDIA RTX~6000 Ada Generation GPU under PyTorch~2.7.1 with CUDA~11.8 enabled.
\subsubsection{Mask Threshold and Guidance Offset Determination}


\begin{table}[t]
\centering
\caption{Mask threshold and guidance offset analysis of SemVideoSte at an SNR of 10~dB.}
\label{tab:parameter_analysis}
\footnotesize
\setlength{\tabcolsep}{2.8pt}
\renewcommand{\arraystretch}{1.08}
\begin{tabular}{@{}
>{\centering\arraybackslash}m{0.15\columnwidth}
>{\centering\arraybackslash}m{0.10\columnwidth}
>{\centering\arraybackslash}m{0.13\columnwidth}
>{\centering\arraybackslash}m{0.16\columnwidth}
>{\centering\arraybackslash}m{0.15\columnwidth}
>{\centering\arraybackslash}m{0.11\columnwidth}
@{}}
\toprule
Parameter & Value & Acc. (\%) & Cap. (bits) & PSNR (dB) & SSIM \\
\midrule

\multirow{5}{*}{$\tau_{\mathrm{sep}}$}
& 0.99 & 99.94 & 781  & 46.20 & 0.9975 \\
& 0.98 & 99.80 & 1603 & 44.18 & 0.9959 \\
& \textbf{0.97} & \textbf{99.72} & \textbf{2400}
& \textbf{43.15} & \textbf{0.9947} \\
& 0.96 & 99.60 & 3203 & 42.47 & 0.9937 \\
& 0.95 & 99.58 & 3995 & 41.98 & 0.9930 \\
\midrule

\multirow{5}{*}{$\tau_{\mathrm{sem}}$}
& 0.02 & 99.80 & 603  & 48.98 & 0.9986 \\
& 0.04 & 99.75 & 1214 & 46.05 & 0.9972 \\
& 0.06 & 99.73 & 1810 & 44.36 & 0.9959 \\
& \textbf{0.08} & \textbf{99.72} & \textbf{2400}
& \textbf{43.15} & \textbf{0.9947} \\
& 0.10 & 99.72 & 3004 & 42.21 & 0.9935 \\
\midrule

\multirow{4}{*}{$\tau_{\mathrm{vae}}$}
& 0.24 & 99.69 & 1851 & 44.17 & 0.9958 \\
& 0.28 & 99.75 & 2127 & 43.62 & 0.9952 \\
& \textbf{0.32} & \textbf{99.72} & \textbf{2400}
& \textbf{43.15} & \textbf{0.9947} \\
& 0.36 & 99.66 & 2655 & 42.76 & 0.9942 \\
\midrule

\multirow{5}{*}{$\Delta cfg_k$}
& 24 & 98.56 & 2396 & 48.76 & 0.9983 \\
& 36 & 99.42 & 2385 & 45.47 & 0.9966 \\
& \textbf{48} & \textbf{99.72} & \textbf{2400}
& \textbf{43.15} & \textbf{0.9947} \\
& 56 & 99.79 & 2408 & 41.92 & 0.9934 \\
& 64 & 99.87 & 2402 & 40.86 & 0.9920 \\
\bottomrule
\end{tabular}
\vspace{-10pt}
\end{table}

To determine the three mask thresholds and the guidance offset, we conduct preliminary experiments on eight randomly selected prompts at an SNR of 10~dB. In each experiment, only one parameter is varied while the others remain unchanged. As shown in Table~\ref{tab:parameter_analysis}, recovery accuracy, embedding capacity, PSNR, and SSIM are jointly evaluated to determine the parameter settings.

The results show that decreasing $\tau_{\mathrm{sep}}$ or increasing $\tau_{\mathrm{sem}}$ and $\tau_{\mathrm{vae}}$ preserves more latent positions for message embedding, thereby increasing the embedding capacity. However, the additionally selected positions may exhibit weaker separability or robustness, resulting in reduced recovery accuracy and video quality. Accordingly, the three mask thresholds are set to $\tau_{\mathrm{sep}}=0.97$, $\tau_{\mathrm{sem}}=0.08$, and $\tau_{\mathrm{vae}}=0.32$. With these thresholds fixed, increasing $\Delta cfg_k$ enhances the distinction between the positive and negative latent states, while an excessively large guidance offset introduces additional video distortion with only marginal improvement in recovery accuracy. Therefore, $\Delta cfg_k=48$ is selected. Under this setting, SemVideoSte achieves a recovery accuracy of 99.72\%, an average capacity of 2400 bits, a PSNR of 43.15~dB, and an SSIM of 0.9947.

\subsection{Evaluation Metrics}
To comprehensively evaluate the performance of SemVideoSte, we employ PSNR, MSE, SSIM, and LPIPS to measure the visual distortion between the original video and the transmitted stego video. PSNR and MSE are adopted to quantify the pixel-level reconstruction distortion, SSIM is used to evaluate the structural similarity, and LPIPS\cite{zhang2018unreasonable} is employed to measure the perceptual difference in deep feature space. Moreover, BERTScore F1 \cite{zhang2019bertscore} is used to evaluate the semantic consistency between the original secret message and the recovered message.

For message recovery, we use recovery accuracy to evaluate whether the secret message embedding can be correctly extracted. Given the original bit sequence $\mathbf{b}=\{b_1,b_2,\ldots,b_L\}$ and the recovered bit sequence $\hat{\mathbf{b}}=\{\hat{b}_1,\hat{b}_2,\ldots,\hat{b}_L\}$, the recovery accuracy is calculated as
\begin{equation}
    \mathrm{Acc}
    =
    \frac{1}{L}
    \sum_{l=1}^{L}
    \mathbb{I}
    \left(
    b_l=\hat{b}_l
    \right),
\end{equation}
where $L$ denotes the length of the secret bit sequence and $\mathbb{I}(\cdot)$ is the indicator function.

For steganalysis evaluation, we adopt the detection error probability $P_E$ to measure whether the steganalyzer can distinguish stego videos from normal videos. It is calculated as
\begin{equation}
    P_E
    =
    \frac{N_{\rm error}}{N_{\rm total}},
\end{equation}
where $N_{\rm error}$ denotes the number of incorrectly classified samples and $N_{\rm total}$ denotes the total number of tested samples. A larger $P_E$ indicates that the steganalyzer has more difficulty in detecting the existence of hidden messages, which means better covertness of the proposed scheme.
\subsection{Simulation Results}

\subsubsection {Performance of the Proposed SemVideoSte Scheme}


\begin{figure*}
\centering
\includegraphics[width=0.9\linewidth]{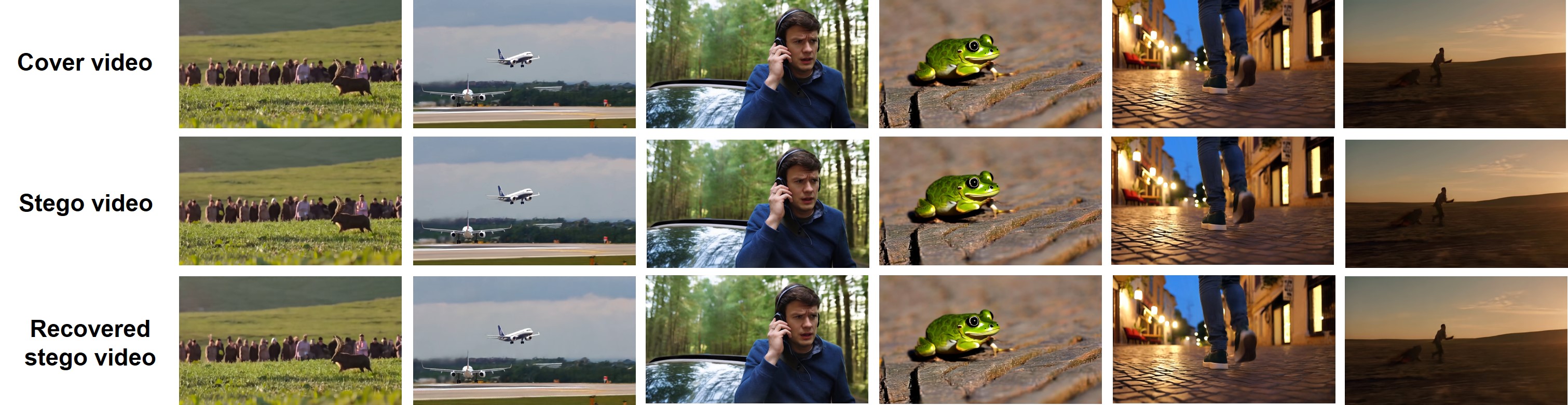}
\caption{Visualization results of SemVideoSte under different public keys. The images are collected at $\text{SNR = 10~dB}$.}
\label{different_public}
\vspace{-10pt}
\end{figure*}

\begin{table}[t]
\centering
\caption{Performance of SemVideoSte under different SNRs.}
\label{tab:main_performance}
\footnotesize
\setlength{\tabcolsep}{6pt}
\renewcommand{\arraystretch}{1.15}
\begin{tabular}{@{}cccc@{}}
\toprule
SNR (dB) & Capacity (bits) & Accuracy (\%) & BERTScore \\
\midrule
5  & 2330 & 97.24 & 0.9068 \\
10 & 2270 & 99.15 & 0.9807 \\
15 & 2233 & 99.47 & 0.9979 \\
20 & 2239 & 99.62 & 0.9889 \\
\bottomrule
\vspace{-20pt}
\end{tabular}
\end{table}

To evaluate the effectiveness of the proposed SemVideoSte scheme, we conduct simulations on 100 public key prompts under different SNRs, where each public key is used to generate a stego video and recover the embedded secret message after semantic transmission. The visualization results under different public keys are presented in Figure~\ref{different_public}. As shown in Table~\ref{tab:main_performance}, SemVideoSte maintains an average embedding capacity above 2200 bits under all tested SNRs. The slight variation in capacity is caused by the different numbers of available positions selected by the semantic-based robust mask under different channel conditions. This indicates that the proposed multi-mask mechanism can retain sufficient embedding positions while filtering positions that are vulnerable to semantic transmission distortion. The recovery accuracy increases from 97.24\% at 5~dB to 99.62\% at 20~dB, demonstrating that the selected latent positions can support reliable message recovery over noisy wireless channels. 

To further evaluate the semantic quality of the recovered information, we fix one public key prompt and independently embed 100 different secret messages. Each recovered message is compared with its corresponding original message using BERTScore F1, and the average value over the 100 messages is reported. As shown in Table~\ref{tab:main_performance}, the average BERTScore F1 remains above 0.90 under all tested SNRs. Even at an SNR of 5~dB, the recovered messages retain their main semantic information despite a small number of incorrectly recovered bits. As the SNR increases, the BERTScore F1 generally reaches a higher level, verifying the semantic consistency between the original and recovered messages.

\begin{table}[t]
\centering
\caption{Video quality of transmitted stego videos under different SNRs.}
\label{tab:transmitted_video_quality}
\footnotesize
\setlength{\tabcolsep}{4.5pt}
\renewcommand{\arraystretch}{1.15}
\begin{tabular}{@{}ccccc@{}}
\toprule
SNR (dB) & PSNR (dB) & SSIM & MSE ($\times 10^{-4}$) & LPIPS \\
\midrule
5  & 32.06 & 0.9116 & 7.876 & 0.1158 \\
10 & 34.21 & 0.9476 & 4.666 & 0.0600 \\
15 & 34.73 & 0.9530 & 4.113 & 0.0605 \\
20 & 34.89 & 0.9542 & 3.965 & 0.0611 \\
\bottomrule
\vspace{-20pt}
\end{tabular}
\end{table}

\begin{figure}
\centering
\includegraphics[width=0.8\linewidth]{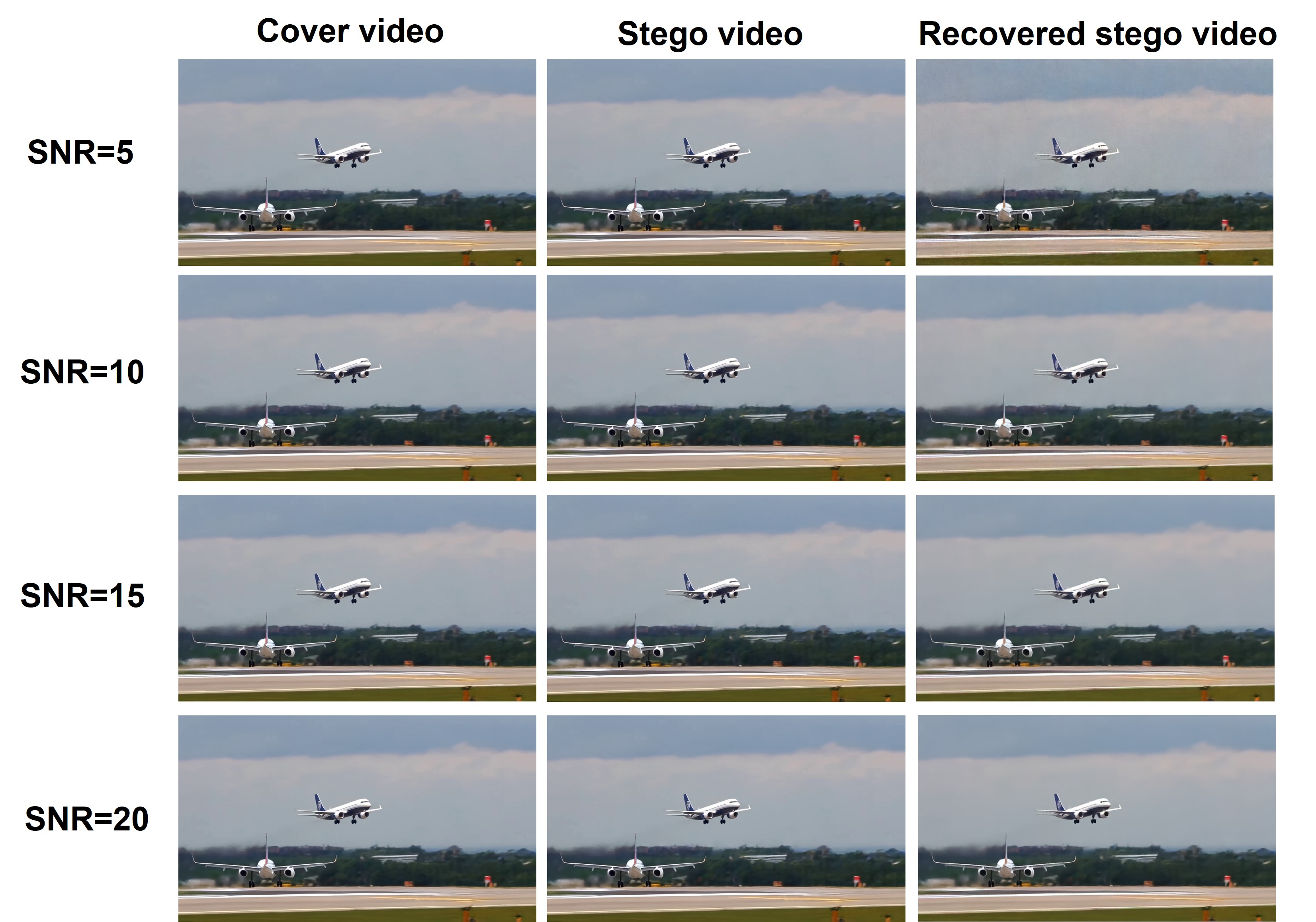}
\caption{Visualization results of SemVideoSte under different SNRs. The public key is ``A plane taxiing down the runway, taking off into the sky and flying over clouds and a scenic landscape.''}
\label{Fig_differentSNR}
\vspace{-10pt}
\end{figure}

Table~\ref{tab:transmitted_video_quality} compares the original videos with the transmitted stego videos in terms of PSNR, SSIM, MSE, and LPIPS. At an SNR of 5~dB, the transmitted stego videos achieve a PSNR of 32.06~dB and an SSIM of 0.9116, indicating that the main visual content and structural information remain well preserved under severe channel noise. As the SNR increases to 20~dB, the PSNR increases to 34.89~dB, while the MSE decreases from $7.876\times10^{-4}$ to $3.965\times10^{-4}$. Meanwhile, the LPIPS remains at a low level under all tested SNRs, demonstrating a small perceptual difference between the original and transmitted stego videos. The visualization results in Figure~\ref{Fig_differentSNR} further show that there is no evident artifacts caused by message embedding. Although semantic transmission introduces slight blurring and texture degradation at low SNRs, the scene content and main object structures remain consistent with the original videos. These results demonstrate that SemVideoSte can embed secret messages during diffusion-based video generation while maintaining satisfactory video quality after semantic transmission.

\begin{figure}
\centering
\includegraphics[width=0.8\linewidth]{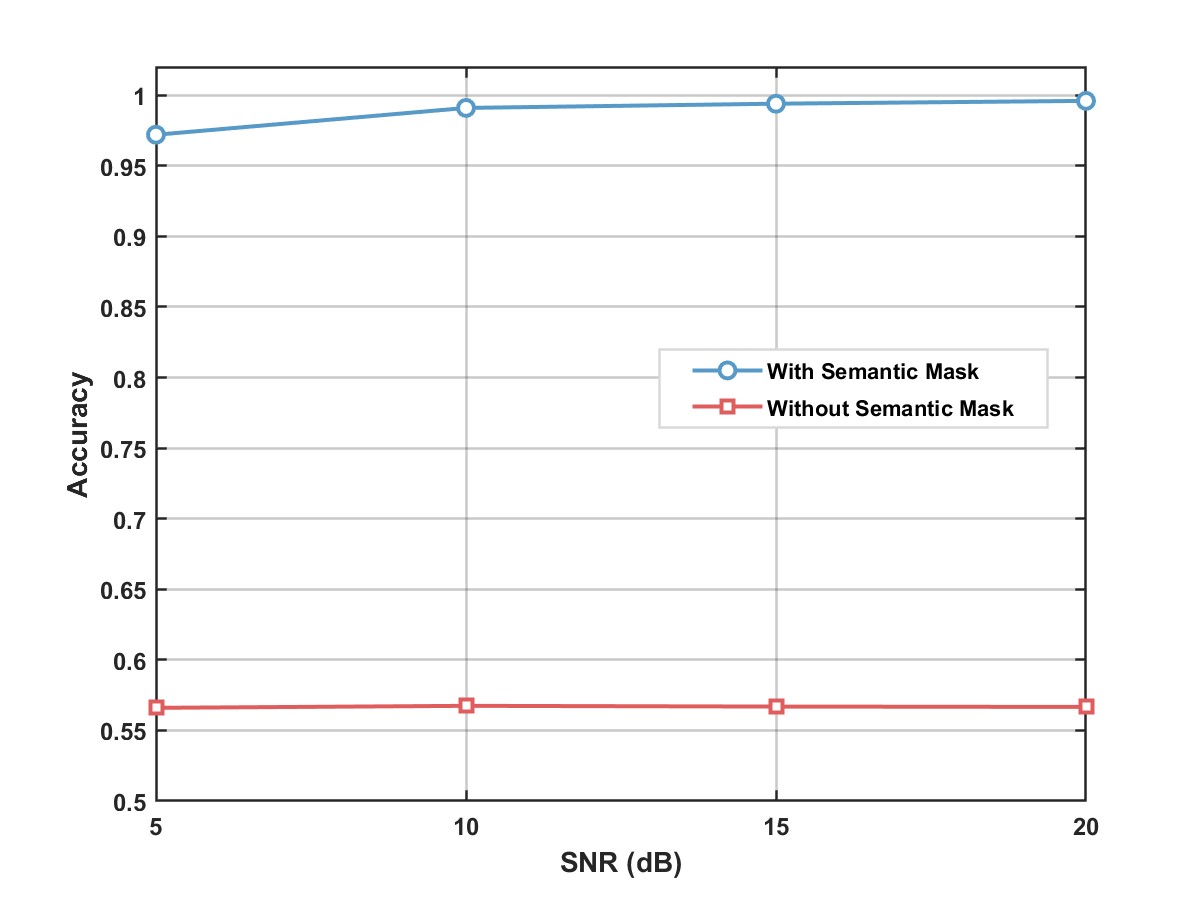}
\caption{Comparison of recovery accuracy between SemVideoSte with and without the semantic-based robust mask.}
\label{fig_without_semantic_mask}
\vspace{-10pt}
\end{figure}


Figure~\ref{fig_without_semantic_mask} compares the proposed recovery accuracy of SemVideoSte with and without the semantic-based robust mask. The proposed scheme achieves a recovery accuracy above 97\% under all tested SNRs, whereas the scheme without this mask maintains an accuracy of only approximately 56.6\%. Although the separation mask selects latent positions with distinguishable positive and negative responses during diffusion sampling, these positions are not necessarily robust to semantic encoding, channel noise, and semantic decoding. Consequently, their latent values may be significantly changed after semantic transmission, resulting in unreliable message decisions at the receiver. The performance difference demonstrates that the semantic-based robust mask is essential for selecting embedding positions that can survive lossy semantic video transmission.

\subsubsection{Security Against Eavesdropping Recovery}
\begin{figure}
\centering
\includegraphics[width=0.8\linewidth]{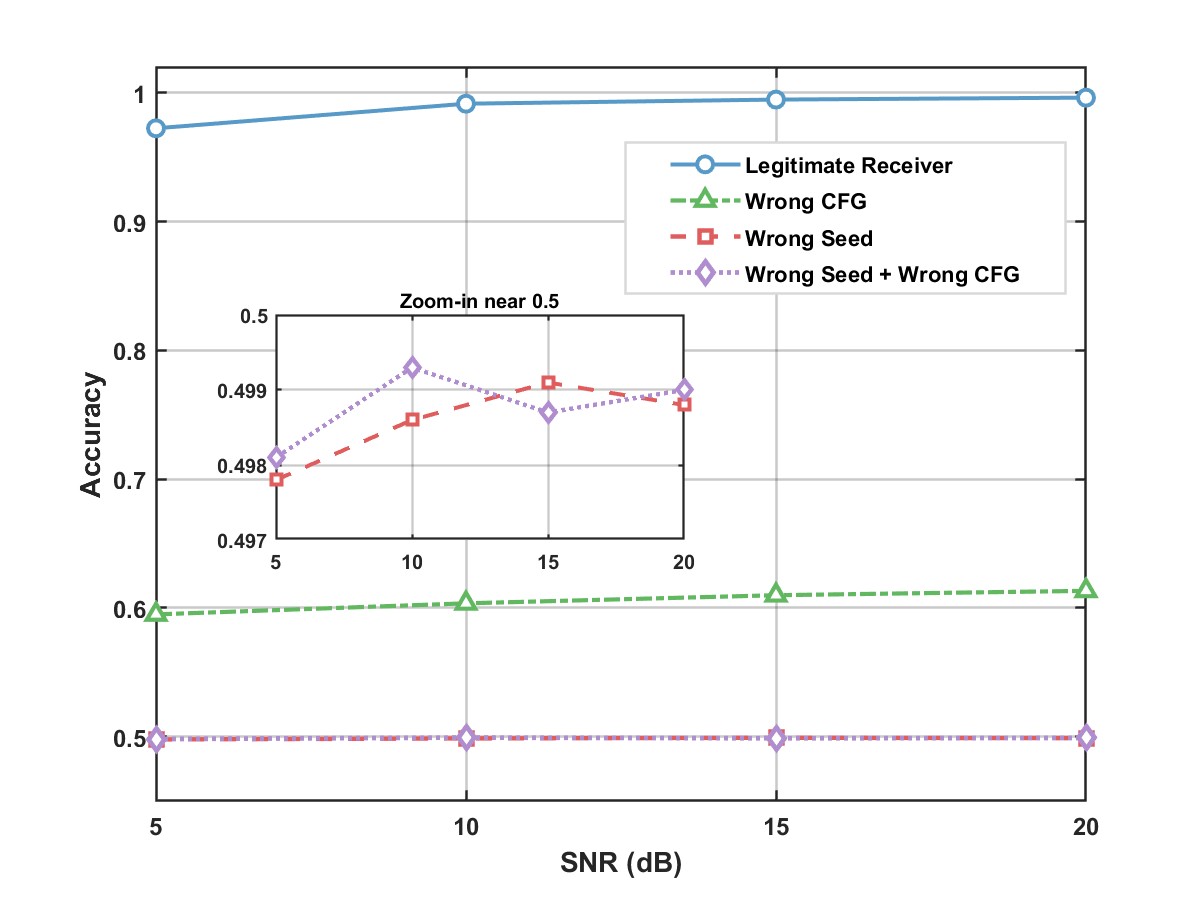}
\caption{Comparison of recovery accuracy between the legitimate receiver and the agentic AI-based eavesdropper, where wrong $\Delta cfg_k$, wrong seed and both wrong parameters are considered.}
\label{Fig_wrongParameter}
\vspace{-5pt}
\end{figure}

To further assess the security performance of SemVideoSte, we consider the worst-case scenario for the agentic AI-based eavesdropper that intercepts the transmitted semantic representation and knows the implementation logic of the proposed scheme \cite{meng2026secure}. The eavesdropper is assumed to possess the public generation model, semantic decoder, and public key $K_{\rm pub}$, but cannot obtain the private digital token $k$, which determines the random seed $s$ and guidance offset $\Delta cfg_k$.
Figure~\ref{Fig_wrongParameter} compares the recovery accuracy of the legitimate receiver with that of the eavesdropper using an incorrect $\Delta cfg_k$, an incorrect seed, or both incorrect parameters. With an incorrect $\Delta cfg_k$, the recovery accuracy is approximately 60\%. Although the correct seed preserves part of the original sampling trajectory, the resulting positive and negative latent references differ from those used for message embedding. When the seed is incorrect, both the sampling trajectory and embedding positions are changed, and the recovery accuracy decreases to approximately 50\%. Using incorrect values for both parameters produces a similar result. This eavesdropping accuracy does not benefit significantly from
increasing SNR, since the improved channel quality  cannot correct the mismatch in the latent references and embedding positions.

Table~\ref{tab:security_bertscore} further compares the BERTScore F1 in eavesdropping cases. The legitimate receiver maintains high semantic consistency, with BERTScore F1 above 0.90 under all SNRs. In contrast, the eavesdroppers' results are much lower. Although the BERTScore of wrong-seed cases remains around 0.75, the recovered text is almost entirely garbled, for the decoding process preserves part of the sentence length and token pattern after bit errors, and BERTScore F1 can assign a moderate score to these residual embedding similarities.  For the wrong $\Delta cfg_k$ case, the BERTScore F1 is slightly higher because several local word fragments are decoded, but they are too incomplete to reveal the precise meaning of the secret message. This further confirms that the eavesdropper cannot obtain useful semantic information without the correct digital token. These results demonstrate that the security of SemVideoSte comes from the token-controlled secure mechanism. Even if the eavesdropper knows the public key and the implementation logic, it can only reconstruct the visible stego video from the intercepted semantic representation, but cannot reproduce the correct latent trajectory and embedding mask. Therefore, the secret message remains hidden in the diffusion-generated video and cannot be effectively extracted without the private digital token.

\begin{table}[t]
\centering
\caption{BERTScore F1 comparison between the legitimate receiver and eavesdroppers under different SNRs.}
\label{tab:security_bertscore}
\footnotesize
\setlength{\tabcolsep}{4.5pt}
\renewcommand{\arraystretch}{1.15}
\begin{tabular}{@{}ccccc@{}}
\toprule
SNR (dB) & Legit. & Wrong $\Delta cfg_k$ & Wrong Seed & Wrong Seed \& $\Delta cfg_k$ \\
\midrule
5  & 0.9068 & 0.8100 & 0.7621 & 0.7510 \\
10 & 0.9807 & 0.8350 & 0.7640 & 0.7530 \\
15 & 0.9979 & 0.8520 & 0.7660 & 0.7560 \\
20 & 0.9889 & 0.8480 & 0.7650 & 0.7540 \\
\bottomrule
\vspace{-20pt}
\end{tabular}
\end{table}

\subsubsection{Steganalysis Evaluation}
\begin{table}[t]
\centering
\caption{Steganalysis detection error of different steganalyzers for SemVideoSte.}
\label{tab:steganalysis_error}
\renewcommand{\arraystretch}{1.25}
\begin{tabular}{cc}
\toprule
\textbf{Steganalyzer} & \textbf{Detection Error $P_E$ (\%)} \\
\midrule
LWENet \cite{weng2022lightweight} & 49.96 \\
XuNet \cite{xu2016structural}  & 50.00 \\
SRNet \cite{boroumand2018deep}  & 49.93 \\
CovNet \cite{deng2019fast} & 38.16 \\
\bottomrule
\end{tabular}
\vspace{-5pt}
\end{table}

Despite preventing the eavesdropper from recovering the secret message, SemVideoSte should also conceal the existence of message embedding. Therefore, we evaluate whether the generated stego videos can be distinguished from normal generated videos by representative steganalyzers. The dataset consists of 100 cover-stego video pairs, with each video containing 81 frames, resulting in 8100 cover-stego frame pairs. Among them, 4000 pairs are used for training, 600 pairs for validation, and 3500 pairs for testing. The detection error $P_E$ is used for evaluation, where a value close to 50\% corresponds to detection performance close to random guessing.

As shown in Table~\ref{tab:steganalysis_error}, four representative steganalyzers with different feature extraction structures are considered. LWENet focuses on lightweight residual feature extraction with multi-view pooling \cite{weng2022lightweight}, XuNet represents a classical convolutional neural network (CNN)-based steganalysis structure with high-pass residual modeling \cite{xu2016structural}, SRNet adopts a deeper residual network to learn subtle stego noise \cite{boroumand2018deep}, and CovNet further exploits covariance pooling to capture second-order residual correlations \cite{deng2019fast}. The detection errors of LWENet, XuNet, SRNet, and CovNet are 49.96\%, 50.00\%, 49.93\%, and 38.16\%, respectively. It can be observed that LWENet, XuNet, and SRNet achieve detection errors very close to 50\%, indicating that these steganalyzers can hardly distinguish the generated cover frames from the stego frames. CovNet achieves a lower detection error, suggesting that it can capture some weak statistical traces caused by the generative embedding process. Nevertheless, the detection error of CovNet is still not sufficiently low to indicate reliable steganalysis, and the generated stego videos remain difficult to distinguish in most cases. Overall, these steganalysis results confirm the covertness of SemVideoSte. By embedding the secret message into mask-selected latent positions during the diffusion generation process, SemVideoSte preserves the natural visual semantics of generated videos and avoids obvious discrepancies that can be captured by deep steganalyzers.

\subsection{Validation Testbed}
\begin{figure}[tbp]
    \centering
    \includegraphics[width=0.8\linewidth]{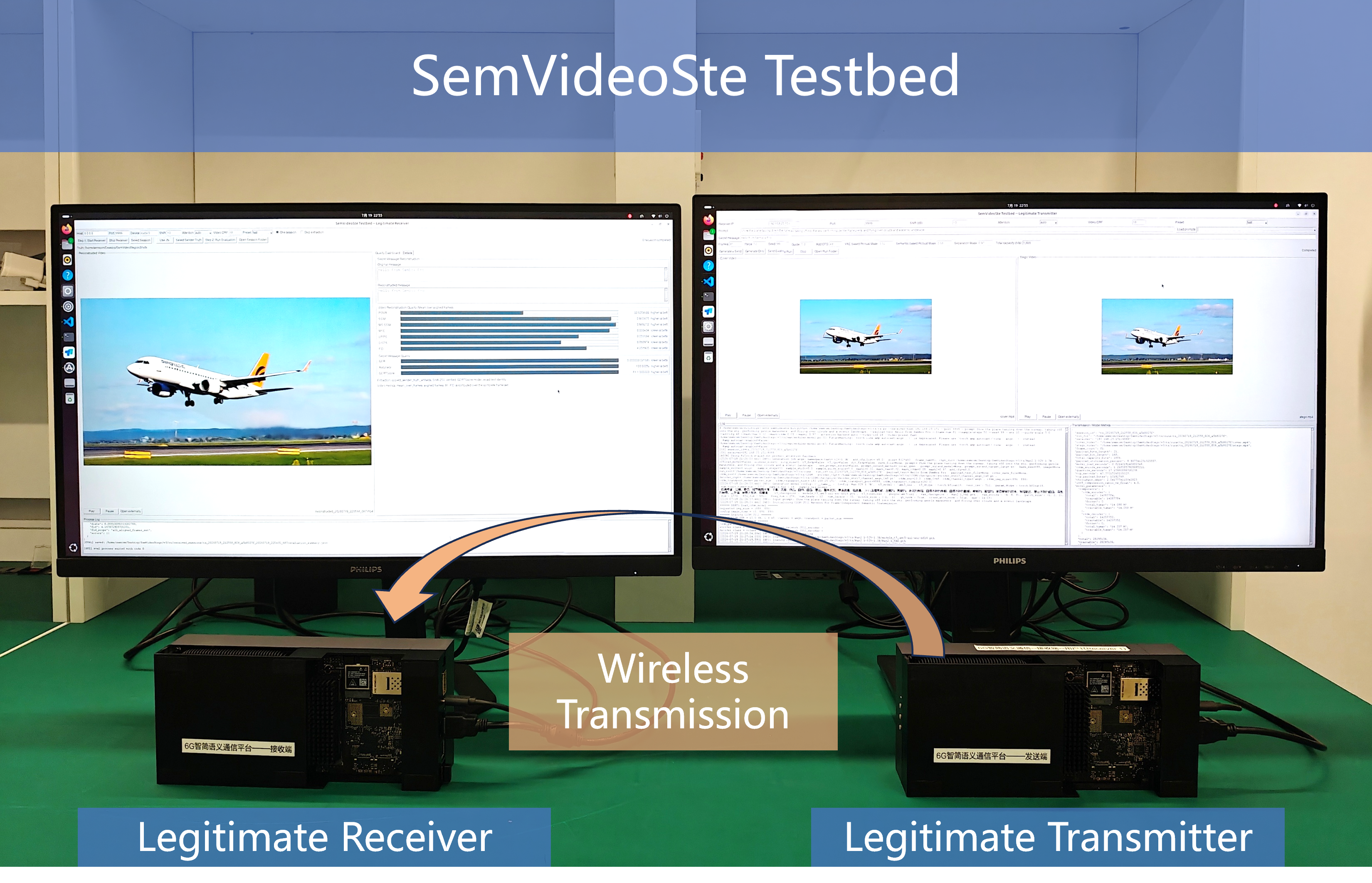}
    \caption{Physical prototype of the developed SemVideoSte validation testbed.}
    \label{testbed}
\vspace{-15pt}
\end{figure}

To evaluate the practical deployability of the proposed SemVideoSte scheme beyond offline simulations, we develop a prototype validation platform, referred to as the SemVideoSte testbed, for end-to-end semantic video steganography communication between two interconnected semantic transceivers. As illustrated in Fig.~\ref{testbed}, the testbed comprises an agentic AI-enabled transmitter and an agentic AI-enabled receiver, each implemented on an NVIDIA Jetson AGX Thor Developer Kit. The two edge devices serve as the legitimate transmitter (Tx) and legitimate receiver (Rx), respectively, enabling the deployment and validation of SemVideoSte on resource-constrained edge platforms. 
For consistency, the testbed experiments employ the same system configurations as the offline simulations, with the detailed parameter settings provided in Sec.~\ref{args}.

\begin{figure}
    \centering
    \begin{subfigure}[b]{0.95\linewidth}
        \centering
        \includegraphics[width=\linewidth]{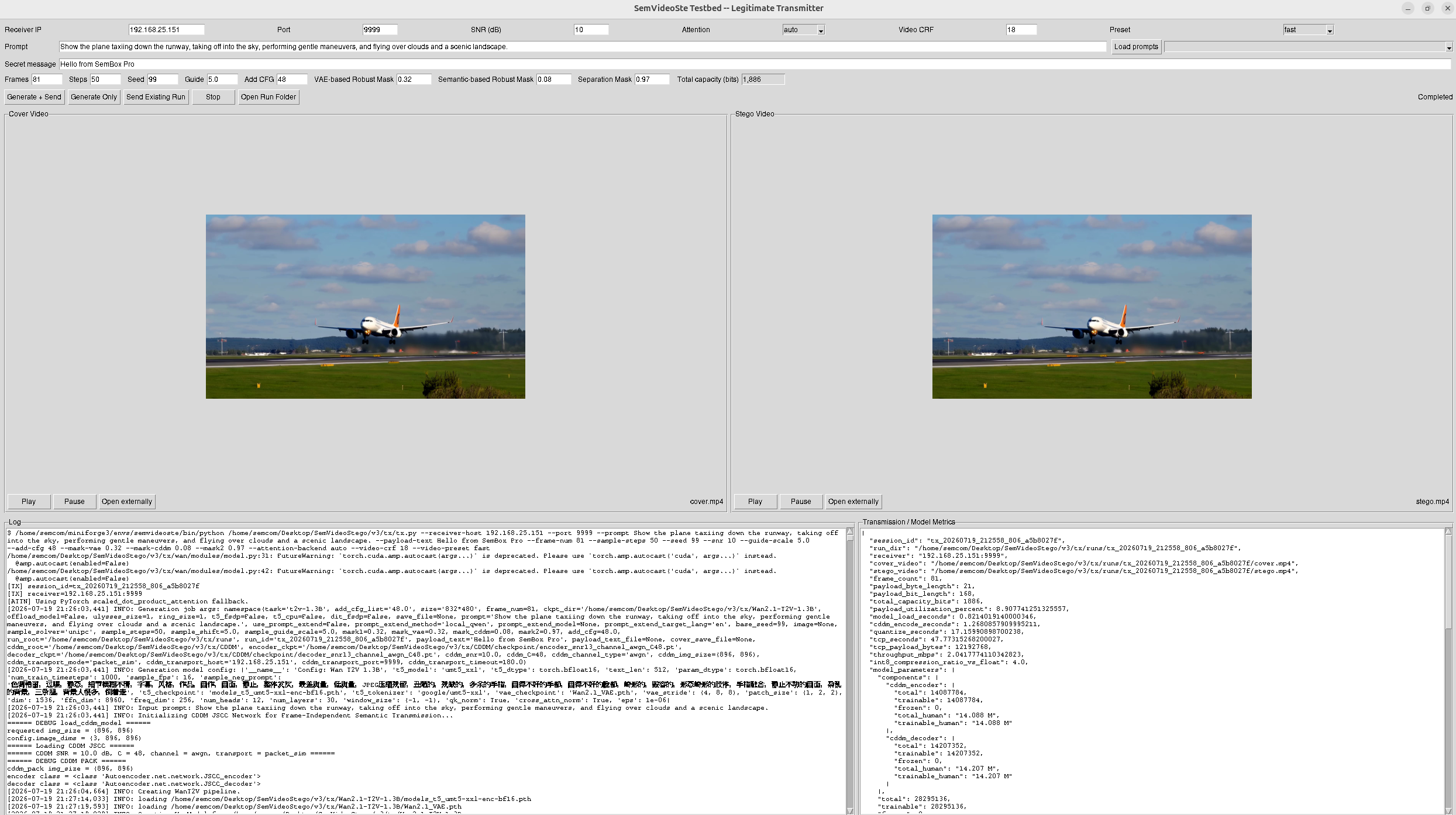}
        \caption{Legitimate transmitter-side interface for semantic video generation, secret embedding, and semantic feature transmission.}
        \label{tx}
    \end{subfigure}
    \par\medskip
    \begin{subfigure}[b]{0.95\linewidth}
        \centering
        \includegraphics[width=\linewidth]{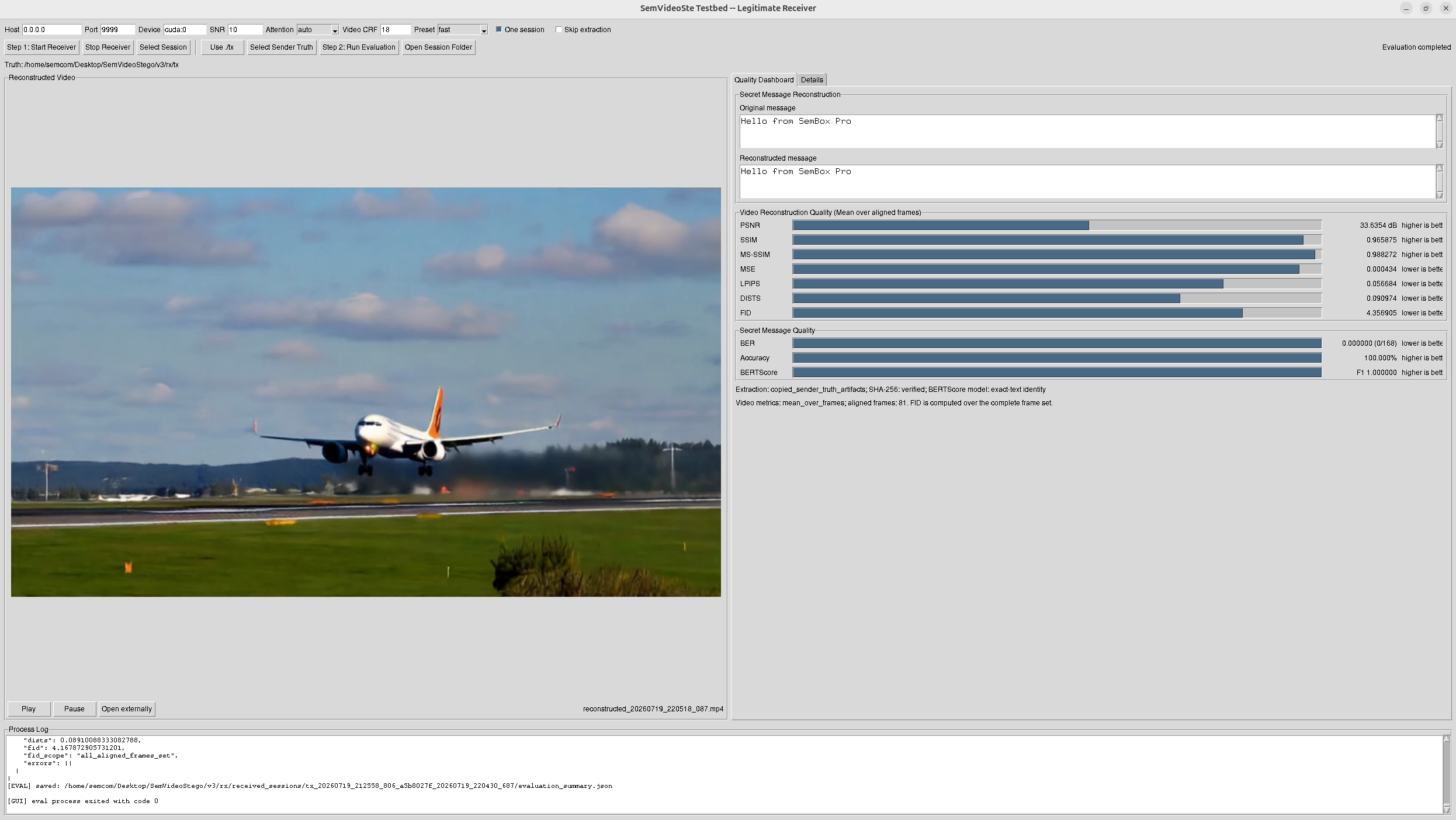}
        \caption{Legitimate receiver-side interface for semantic video reconstruction, secret message recovery, and performance evaluation.}
        \label{rx}
    \end{subfigure}
    \caption{Dedicated transmitter-side and receiver-side software interfaces of the developed SemVideoSte testbed.}
    \label{testbed-visual}
\end{figure}

At the transmitter, as shown in Fig.~\ref{tx}, the agentic AI-enabled transceiver first generates a cover video from a public prompt using the Wan2.1 text-to-video diffusion model. The secret message is subsequently embedded into the latent semantic representation of the generated video to produce a stego video, which is then encoded by the JSCC-based semantic codec and quantized into a compact semantic feature representation. The resulting semantic features are delivered to the receiver over the wireless communication link using TCP/IP as the transport protocol. At the receiver, as shown in Fig.~\ref{rx}, the received semantic features are decoded to reconstruct the transmitted video sequence. The receiver further regenerates the reference latent representations using the same public prompt and diffusion parameters as those employed at the transmitter, and then performs semantic extraction to recover the embedded secret message. The resulting end-to-end procedure follows a practical digital SemCom pipeline while retaining the coverless steganographic mechanism of the proposed SemVideoSte scheme.

To facilitate system configuration, inference monitoring, and experimental analysis, dedicated software interfaces are developed for both the legitimate transmitter and receiver, as shown in Fig.~\ref{testbed-visual}. The transmitter-side interface supports the configuration of the public prompt, secret message, video generation parameters, SemCom parameters, three mask thresholds, embedding capacity, and diffusion hyperparameters. During system operation, the generated cover and stego videos are displayed simultaneously, together with transmission logs, runtime statistics, and model information. The receiver-side interface visualizes the reconstructed video, recovered secret message, communication status, and quantitative evaluation results. In addition to inference monitoring, the receiver records the transmission latency, semantic decoding throughput, model complexity, and video reconstruction statistics for subsequent performance analysis.

The developed testbed also incorporates an automated evaluation module for assessing both semantic steganography and SemCom performance. For secret message recovery, the bit error rate (BER) and BERTScore F1 are reported to characterize bit-level recovery accuracy and semantic consistency, respectively. For video reconstruction, complementary pixel-level, structural-level, and perceptual-level quality metrics are employed, including PSNR, SSIM, MS-SSIM, LPIPS, DISTS, and FID. These metrics are computed by comparing the reconstructed video at the receiver with the corresponding reference video available at the transmitter. Therefore, the developed testbed enables both qualitative visualization and quantitative evaluation of the proposed SemVideoSte scheme, thereby demonstrating its feasibility and effectiveness in a practical deployment environment.

\section{Conclusions}

In this paper, we have proposed the SemVideoSte scheme for hiding confidential textual auxiliary information in semantic video services. Based on the service context and transmission requirements, agentic AI determines a suitable public prompt and organizes the subsequent video generation, message embedding, semantic transmission, and recovery processes. The secret textual message is embedded into the generative latent space during the conditional diffusion process, where the multi-mask mechanism selects positions that are robust to reconstruction and semantic transmission distortion and exhibit distinguishable guidance responses. In this manner, the public video maintains its normal visual semantics, while the confidential text is transmitted through a cross-modal steganographic path without requiring a pre-existing video carrier. The private digital token further determines the random seed and guidance offset used to construct the latent references and embedding mask. 
Without the correct token, an eavesdropper cannot reproduce the corresponding sampling trajectory and recovery references, which effectively prevents unauthorized extraction of confidential messages and strengthens the security of SemVideoSte against intelligent eavesdropping.
Experimental results demonstrate the effectiveness of SemVideoSte in terms of embedding capacity, message recovery reliability, semantic consistency, video quality, and resistance to eavesdropping and steganalysis. The developed SemVideoSte testbed further verifies the feasibility of our scheme on edge platforms.

\bibliographystyle{IEEEtran}
\bibliography{ref}

@article{huo2025image,
  title={Image Semantic Steganography: A Way to Hide Information in Semantic Communication},
  author={Huo, Yanhao and others},
  journal={IEEE Transactions on Circuits and Systems for Video Technology},
  year={2025},
  volume={35},
  number={2},
  pages={1951-1960},
  publisher={IEEE}
}

@article{tang2025towards,
  title={Towards Secure Semantic Communications in the Presence of Intelligent Eavesdroppers},
  author={Tang, Shunpu and others},
  journal={arXiv preprint arXiv:2503.23103},
  year={2025}
}

@article{zhang2026towards,
  title={Towards Semantic-based Agent Communication Networks: Vision, Technologies, and Challenges},
  author={Zhang, Ping and others},
  journal={SCIENCE CHINA Information Sciences},
  year={2026}
}

@article{meng2025semantic,
      title={Semantic Radio Access Networks: Architecture, State-of-the-Art, and Future Directions}, 
      author={Rui Meng and others},
  journal={IEEE Transactions on Cognitive Communications and Networking},
  volume={12},
  pages={7076--7097},
  year={2026},
  publisher={IEEE}
}

@article{meng2025survey,
  title={A survey of secure semantic communications},
  author={Meng, Rui and others},
  journal={Journal of Network and Computer Applications},
  volume={239},
  pages={104181},
  year={2025},
  publisher={Elsevier}
}

@article{meng2025secure,
  title={Secure semantic communication with homomorphic encryption},
  author={Meng, Rui and others},
  journal={arXiv preprint arXiv:2501.10182},
  year={2025}
}

@article{meng2026intellicise,
  title={Intellicise Wireless Networks Meet Agentic AI: A Security and Privacy Perspective},
  author={Meng, Rui and others},
  journal={arXiv preprint arXiv:2602.15290},
  year={2026}
}

@article{chencoding,
  author={Chen, Weixuan and others},
  journal={IEEE Journal on Selected Areas in Communications}, 
  title={Can Knowledge Improve Security? A Coding-Enhanced Jamming Approach for Semantic Communication}, 
  year={2026},
  volume={44},
  number={},
  pages={1765-1780},
}

@article{kingma2013auto,
  title={Auto-encoding variational bayes},
  author={Kingma, Diederik P and Welling, Max},
  journal={arXiv preprint arXiv:1312.6114},
  year={2013}
}

@article{ho2022classifier,
  title={Classifier-free diffusion guidance},
  author={Ho, Jonathan and others},
  journal={arXiv preprint arXiv:2207.12598},
  year={2022}
}

@article{yang2024swinjscc,
  title={Swinjscc: Taming swin transformer for deep joint source-channel coding},
  author={Yang, Ke and others},
  journal={IEEE Transactions on Cognitive Communications and Networking},
  volume={11},
  number={1},
  pages={90--104},
  year={2024},
  publisher={IEEE}
}

@inproceedings{zhang2018unreasonable,
  title={The unreasonable effectiveness of deep features as a perceptual metric},
  author={Zhang, Richard and others},
  booktitle={Proceedings of the IEEE conference on computer vision and pattern recognition},
  pages={586--595},
  year={2018}
}

@inproceedings{tung2023deep,
  title={Deep joint source-channel and encryption coding: Secure semantic communications},
  author={Tung, Tze-Yang and others},
  booktitle={ICC 2023-IEEE International Conference on Communications},
  pages={5620--5625},
  year={2023},
  organization={IEEE}
}

@inproceedings{xu2023latent,
  title={Latent semantic diffusion-based channel adaptive de-noising SemCom for future 6G systems},
  author={Xu, Bingxuan and others},
  booktitle={GLOBECOM 2023-2023 IEEE Global Communications Conference},
  pages={1229--1234},
  year={2023},
  organization={IEEE}
}

@article{ni2025controllable,
  title={Controllable Steganography for Robust and Efficient Semantic Communication Systems},
  author={Ni, Shouxiang and others},
  journal={IEEE Communications Magazine},
  year={2026},
  volume={64},
  number={1},
  pages={160-166},
  publisher={IEEE}
}

@article{gao2026semstediff,
  title={Semstediff: Generative diffusion model-based coverless semantic steganography communication},
  author={Gao, Song and others},
  journal={IEEE Transactions on Mobile Computing},
  year={2026},
  publisher={IEEE}
}

@article{jiang2025large,
  title={From large ai models to agentic ai: A tutorial on future intelligent communications},
  author={Jiang, Feibo and others},
  journal={IEEE Journal on Selected Areas in Communications}, 
  year={2026},
  volume={44},
  number={},
  pages={3507-3540},
}

@article{karras2017progressive,
  title={Progressive growing of gans for improved quality, stability, and variation},
  author={Karras, Tero and others},
  booktitle={International Conference on Learning Representations},
  year={2018}
}

@article{gao2025agentic,
  title={Agentic AI-Enhanced Semantic Communications: Foundations, Architecture, and Applications},
  author={Gao, Haixiao and others},
  journal={arXiv preprint arXiv:2512.23294},
  year={2025}
}

@article{zhang2024interactive,
  title={Interactive AI with retrieval-augmented generation for next generation networking},
  author={Zhang, Ruichen and others},
  journal={IEEE Network},
  volume={38},
  number={6},
  pages={414--424},
  year={2024},
  publisher={IEEE}
}

@article{zhang2024experts,
  title={Generative AI agents with large language model for satellite networks via a mixture of experts transmission},
  author={Zhang, Ruichen and others},
  journal={IEEE Journal on Selected Areas in Communications},
  year={2024},
  volume={42},
  number={12},
  pages={3581-3596},
  publisher={IEEE}
}

@article{zhang2025toward,
  title={Toward agentic ai: generative information retrieval inspired intelligent communications and networking},
  author={Zhang, Ruichen and others},
  journal={arXiv preprint arXiv:2502.16866},
  year={2025}
}

@article{zhang2024space-air-ground,
  title={Generative AI for space-air-ground integrated networks},
  author={Zhang, Ruichen and others},
  journal={IEEE Wireless Communications},
  volume={31},
  number={6},
  pages={10--20},
  year={2024},
  publisher={IEEE}
}

@article{xu2016structural,
  title={Structural design of convolutional neural networks for steganalysis},
  author={Xu, Guanshuo and others},
  journal={IEEE Signal Processing Letters},
  volume={23},
  number={5},
  pages={708--712},
  year={2016},
  publisher={IEEE}
}

@article{tang2024retrieval,
  title={Retrieval-augmented generation for {GenAI}-enabled semantic communications},
  author={Tang, Shunpu and others},
  journal={{IEEE} Wirel. Commun.},
  year={2026},
  volume={33},
  number={1},
  pages={259-268},
}

@article{garcia2019essential,
  title={The essential guide to realizing 5G-connected UAVs with massive MIMO},
  author={Garcia-Rodriguez, Adrian and others},
  journal={IEEE Communications Magazine},
  volume={57},
  number={12},
  pages={84--90},
  year={2019},
  publisher={IEEE}
}

@article{zeng2019accessing,
  title={Accessing from the sky: A tutorial on UAV communications for 5G and beyond},
  author={Zeng, Yongs and others},
  journal={Proceedings of the IEEE},
  volume={107},
  number={12},
  pages={2327--2375},
  year={2019},
  publisher={IEEE}
}

@article{ko2021drone,
  title={Drone secure communication protocol for future sensitive applications in military zone},
  author={Ko, Yongho  and others},
  journal={Sensors},
  volume={21},
  number={6},
  pages={2057},
  year={2021},
  publisher={MDPI}
}

@inproceedings{bennett2013analysis,
  title={Analysis of operational airborne ISR full motion video metadata},
  author={Bennett, Bruce and others},
  booktitle={MILCOM 2013-2013 IEEE Military Communications Conference},
  pages={323--327},
  year={2013},
  organization={IEEE}
}

@ARTICLE{11535035,
  author={Tian, Mengqiu and others},
  journal={IEEE Transactions on Mobile Computing}, 
  title={Enhancing Service Function Chain Performance in SDN/NFV-Enabled Space-Air-Ground Integrated Vehicular Networks: A Three-Sided Stable Matching Approach}, 
  year={2026},
  volume={},
  number={},
  pages={1-17}
}

@article{meng2025image,
  title={Image Steganography For Securing Intellicise Wireless Networks:" Invisible Encryption" Against Eavesdroppers},
  author={Meng, Rui and others},
  journal={arXiv preprint arXiv:2505.04467},
  year={2025}
}

@article{gao2025adaptive,
  title={Adaptive cross-modal super-resolution semantic communication for mobile AI-generated panoramic video},
  author={Gao, Haixiao and others},
  journal={IEEE Transactions on Cognitive Communications and Networking},
  year={2025},
  publisher={IEEE}
}

@article{kunhoth2023video,
  title={Video steganography: recent advances and challenges},
  author={Kunhoth, Jayakanth and others},
  journal={Multimedia Tools and Applications},
  volume={82},
  number={27},
  pages={41943--41985},
  year={2023},
  publisher={Springer}
}

@inproceedings{mou2023large,
  title={Large-capacity and flexible video steganography via invertible neural network},
  author={Mou, Chong and others},
  booktitle={Proceedings of the IEEE/CVF conference on computer vision and pattern recognition},
  pages={22606--22615},
  year={2023}
}

@article{meng2026secure,
  title={Secure Intellicise Wireless Network: Agentic AI for Coverless Semantic Steganography Communication},
  author={Meng, Rui and others},
  journal={IEEE Transactions on Cognitive Communications and Networking},
  year={2026}
}

@article{fan2026generative,
  title={Generative diffusion models for wireless networks: Fundamental, architecture, and state-of-the-art},
  author={Fan, Dayu and others},
  journal={IEEE Communications Surveys \& Tutorials},
  year={2026},
  publisher={IEEE}
}

@article{wang2026ld,
  title={LD-RoViS: Training-free Robust Video Steganography for Deterministic Latent Diffusion Model},
  author={Wang, Xiangkun and others},
  journal={Advances in Neural Information Processing Systems},
  volume={38},
  pages={128397--128420},
  year={2026}
}

@article{zhang2026toward,
  title={Toward edge general intelligence with agentic AI and agentification: Concepts, technologies, and future directions},
  author={Zhang, Ruichen and others},
  journal={IEEE Communications Surveys \& Tutorials},
  volume={28},
  pages={4285--4318},
  year={2026},
  publisher={IEEE}
}

@article{abou2025agentic,
  title={Agentic AI: a comprehensive survey of architectures, applications, and future directions},
  author={Abou Ali, Mohamad and others},
  journal={Artificial Intelligence Review},
  volume={59},
  number={1},
  pages={11},
  year={2025},
  publisher={Springer}
}

@article{zhang2019bertscore,
  title={Bertscore: Evaluating text generation with bert},
  author={Zhang, Tianyi and others},
  journal={arXiv preprint arXiv:1904.09675},
  year={2019}
}

@article{lipman2022flow,
  title={Flow matching for generative modeling},
  author={Lipman, Yaron and others},
  journal={arXiv preprint arXiv:2210.02747},
  year={2022}
}

@article{tang2026eavesdroppers,
  title={When Eavesdroppers Reason: Agentic Eavesdropping Attacks on Semantic Communication},
  author={Tang, Shunpu and others},
  journal={arXiv preprint arXiv:2605.02204},
  year={2026}
}

@article{tan2021coverless,
  title={Coverless steganography based on motion analysis of video},
  author={Tan, Yun and others},
  journal={Security and Communication Networks},
  volume={2021},
  number={1},
  pages={5554058},
  year={2021},
  publisher={Wiley Online Library}
}

@article{meng2023robust,
  title={A robust coverless video steganography based on the similarity of inter-frames},
  author={Meng, Laijin and others},
  journal={IEEE Transactions on Multimedia},
  volume={26},
  pages={5996--6011},
  year={2023},
  publisher={IEEE}
}

@inproceedings{rombach2022high,
  title={High-resolution image synthesis with latent diffusion models},
  author={Rombach, Robin and others},
  booktitle={Proceedings of the IEEE/CVF conference on computer vision and pattern recognition},
  pages={10684--10695},
  year={2022}
}

@article{dao2023flow,
  title={Flow matching in latent space},
  author={Dao, Quan and others},
  journal={arXiv preprint arXiv:2307.08698},
  year={2023}
}

@article{boroumand2018deep,
  title={Deep residual network for steganalysis of digital images},
  author={Boroumand, Mehdi and others},
  journal={IEEE Transactions on Information Forensics and Security},
  volume={14},
  number={5},
  pages={1181--1193},
  year={2018},
  publisher={IEEE}
}

@article{weng2022lightweight,
  title={Lightweight and effective deep image steganalysis network},
  author={Weng, Shaowei and others},
  journal={IEEE Signal Processing Letters},
  volume={29},
  pages={1888--1892},
  year={2022},
  publisher={IEEE}
}

@inproceedings{deng2019fast,
  title={Fast and effective global covariance pooling network for image steganalysis},
  author={Deng, Xiaoqing and others},
  booktitle={Proceedings of the ACM workshop on information hiding and multimedia security},
  pages={230--234},
  year={2019}
}

\end{document}